\documentclass[preprint,times]{elsarticle}
\usepackage[a4paper,left=1in,right=1in,top=1in,bottom=1in,footskip=.25in]{geometry}

\usepackage{amsmath}
\usepackage{float} 

\usepackage{bm}
\usepackage{enumitem}

\usepackage{commath}
\usepackage{mathtools}
\usepackage{upgreek}
\usepackage{xcolor}
\usepackage{color} 
\newcommand{\rd}[1]{\textcolor[rgb]{0.00,0.00,0.00}{#1}}

\biboptions{numbers,sort&compress}

\journal{Construction and Building Materials}

\begin{document}

\begin{frontmatter}


\title{Self-powered weigh-in-motion system combining vibration energy harvesting and self-sensing composite pavements}

\author[inst1]{Hasan Borke Birgin}
\ead{hasanborke.birgin@unipg.it}
\author[inst2]{Enrique Garc\'{i}a-Mac\'{i}as\corref{cor1}}
\ead{enriquegm@ugr.es}
\author[inst1]{Antonella D'Alessandro}
\ead{antonella.dalessandro@unipg.it}
\author[inst1]{Filippo Ubertini}
\ead{filippo.ubertini@unipg.it}
\address[inst1]{Department of Civil and Environmental Engineering, University of Perugia. Via G. Duranti, 93 - 06125 Perugia, Italy.}
\address[inst2]{Department of Structural Mechanics and Hydraulic Engineering, University of Granada, Av. Fuentenueva sn, 18002 Granada, Spain.}
\cortext[cor1]{Corresponding author. Department of Structural Mechanics and Hydraulic Engineering, University of Granada, Av. Fuentenueva sn, 18002 Granada, Spain. phone: +34 958241000 Ext. 20668 }

\date{}

\begin{abstract}
\rd{Overloaded} vehicles are the primary cause of accelerated degradation \rd{of road infrastructures}. In this context, although weigh-in-motion (WIM) systems are most efficient to enforce weight regulations, current technologies require costly investments limiting their extensive implementation. Recent advances in multifunctional composites enabled cost-efficient alternatives in the form of smart pavements. Nevertheless, the need for a stable power supply still represents a major practical limitation. This work presents a novel proof-of-concept self-sustainable WIM technology combining smart pavements and vibration-based energy harvesting (EH). The feasibility of piezoelectric bimorph cantilevered beams to harvest traffic-induced vibrations is firstly investigated, followed by the demonstration of the proposed technology under laboratory conditions. \rd{The main original contributions of this work comprise (i) the development of a new self-powered data acquisition system, (ii) a novel approach for the fabrication and electromechanical testing of the piezoresistive composite pavement, and (iii) laboratory feasibility analysis of the developed EH unit to conduct traffic load identification through electrical resistivity measurements of the smart pavement}. While the presented results conclude the need for dense EH networks or combinations of different EH technologies to attain complete self-sustainability, this work represents an initial \rd{feasibility evidence paving the way towards the development of} self-powered low-cost WIM systems.  
\end{abstract}


\end{frontmatter}

\section{Introduction}

Over the last few decades, there has been a noticeable increase in the volume of overweight vehicles worldwide. For instance, the EU fleet registered 3.9 million goods vehicles in 2020 (7.3\% more than by 2019), from which heavy vehicles with maximum laden weights between 10.1 to 20 tonnes represented between 70 to 97\% the total road freight transport across Europe~\cite{Eurostat}. This trend has raised growing concern \rd{among} transportation agencies over the contribution of overweight vehicles to \rd{accelerating the} deterioration rates of pavements and bridges as well as the associated increases in maintenance, upgrading and replacement costs of highway infrastructures~\cite{Fiorillo2018}. For instance, 17 bridge collapses (10.83\% the total number of collapses) directly related to overweight vehicles were registered in China since 2000~\cite{Fu2013}. In the U.S., the last Infrastructure Report Card issued by the American Society of Civil Engineers in 2021~\cite{ASCE2021} assigned a grade of ``D'' (poor, at risk) to the condition of the national highways, \rd{with} overloading \rd{representing} the third primary cause of bridge failures after hydraulic events and collisions~\cite{Wardhana2003}. This has motivated the enforcement of stricter maximum weight policies along with the development of efficient vehicle weighing technologies. In this light, WIM technologies complementing or substituting traditional static weighing stations are regarded as an efficient solution to enforce weight regulations while causing minimum disruption to the traffic flow. 

Weigh-in-motion systems allow to weigh vehicles under normal traffic conditions and provide valuable information for infrastructure managers such as traffic volume, vehicle's speed, axle spacing and weight~\cite{Zhang2007}. There exist two main WIM approaches, namely pavement WIM and bridge WIM (B-WIM). Pavement WIM systems represent the most popular approach and consist of sensors embedded in the pavement to measure the instantaneous forces while tires pass by~\cite{Jacob2002}. BWIM systems, instead, exploit the response of bridges (e.g.~displacements, accelerations) as a scale to identify the weights of vehicles passing overhead~\cite{Moses1979}. Nevertheless, such systems require the use of complex data analysis techniques to identify trucks. In particular, the effect of trucks upon the structural response of bridges is affected by numerous factors such as the presence of multiple vehicles, environmental factors, the intrinsic dynamic behaviour of the bridge, and vehicle-structure interaction effects~\cite{OBrien2001}. The basic configuration of pavement WIM systems consists of at least one weight sensor and two inductive loops embedded in a road cut to trigger the sensor when a vehicle passes by~\cite{Burnos2007}, making the data processing considerably simpler. The most commonly used devices for pavement-based WIM systems include bending plates, load cells, capacitance mats, and strip sensors.~\cite{Dontu2020}. Despite the advanced state of development of these technologies, there remain considerable obstacles for their extensive implementation, including difficulties to obtain accurate weight measurements due to the interaction with the vehicle dynamics, challenging vehicle classification, and elevated installation and maintenance costs. These limitations have fostered considerable research efforts to enhance the accuracy and reduce the cost of WIM systems. These include the development of portable WIM systems~\cite{Faruk2016} and computer vision techniques exploiting photos from traffic video surveillance~\cite{Santero2005} or internet protocol (IP)~\cite{He2019} cameras. A comprehensive state-of-the-art review of recent developments, vision and challenges of low-cost WIM systems can be found in reference~\cite{Sujon2021}. 

Recent breakthroughs in the realm of Nanotechnology and Materials Science have enabled the development of new multifunctional composite materials with vast applicability across disciplines. In particular, self-diagnostic materials suggest the development of next-generation smart structures and open broad possibilities in the realm of Structural Health Monitoring (SHM)~\cite{Li2022}. Particularly popular are polymer or cementitious materials doped with carbon-based fillers, including carbon fibers (CFs), carbon black (CB), carbon nanotubes (CNTs), graphene (G) and derivatives such as graphene nano-platelets (GnPs), graphene oxide (GO), or reduced graphene oxide (rGO)~\cite{Tian2019}. The sensing principle of such materials usually lies on their piezoresistive behaviour, in such a way that their strain condition when subjected to mechanical loads can be inferred from variations in their electrical conductivity~\cite{Han2012}. In this light, considerable research efforts have been reported in the literature to optimise the manufacturing process, scalability and electromechanical properties of self-sensing materials. It is worth noting the contribution by D'Alessandro and co-authors~\cite{DAlessandro2016} who investigated the use of different dispersion agents and mixing procedures to optimize the strain self-sensing capabilities of cement-based materials doped with Multi-Walled CNTs (MWCNTs). Their results reported maximum gauge factors (GFs) of 130, 68 and 23 for pastes, mortars and concretes, respectively. \rd{The} study by Suo \textit{et al}.~\cite{Suo2022} reported the electromechanical characterization of cement composites doped with GO at different concentrations. Their results informed increases in the compressive strength with respect to pristine cement from 61.87 to 71.79 MPa (16\%) and GFs in the range of 5-12 for a filler concentration of 0.1 wt.\%. \rd{A recent study by Ding \textit{et al}.~\cite{ding2022situ} reported the fabrication of self-sensing cementitious composites through in-situ synthesis of CNTs on the cement particles, achieving a maximum stress sensitivity of 2.87\%/MPa and a gauge factor of 748.} These results evidence the possibility of developing load-bearing sensors with high durability (in principle equal to the matrix material), high compatibility with the host structure (no ancillaries are required for the installation), and high accuracy since strain measurements are taken directly from the structure experiencing the mechanical loading. This concept has been explored in the literature in the shape of embeddable sensors~\cite{Meoni2018}, smart skins~\cite{Schumacher2014}, and complete structural elements~\cite{Downey2017}. Nevertheless, the development of self-sensing composite materials for SHM of pavements has been scarcely investigated.

Strain measurements through piezoresistive materials are typically performed by DC or AC resistivity measurements between electrodes inserted in or attached to the material~\cite{Han2015,Downey2017a,Bekzhanova2021}. This poses a major limitation for the extensive technological transfer of self-sensing materials to routine engineering practice, since securing stable on-site power supplies during their life-time is often infeasible. This explains the fact that very few long-term field applications have been reported in the literature, being most applications limited to laboratory environments. This is a common limitation in most sensing technologies, and although the use of batteries may represent a viable solution, their limited durability remains a considerable drawback. Solar panels, while capable of providing sufficient power on sunny days for regular sensor data collection and transmission, produce limited or no power during night-times and cloudy periods. In this light, considerable research efforts have been exerted in the last decades to develop efficient EH technologies capable of enabling self-powered SHM systems. In the realm of civil engineering structures, particular attention has been devoted to vibration-based EH systems. The harvesting principle of these systems lies on the conversion of vibration sources such as ocean waves, human motion, wind forces or structural vibrations into electricity~\cite{Wei2017,wang2017magnetic,maamer2019review,wang2019high,ma2022review}. This principle is particularly well-suited for transport infrastructures such as rail~\cite{Qi2022} or road~\cite{Pei2021} ways, where vibrations induced by traffic and environmental loads are considerable and frequent. Piezoelectric harvesters are especially popular due to their high energy conversion efficiency, ease of implementation, and miniaturization. These devices usually integrate different configurations of piezoceramics such as lead zirconate titanate (PZT) or polyvinylidene difluoride (PVDF). Upon the application of stress, these materials generate electric charges by the so-called direct piezoelectric effect~\cite{Yang2018}. 

Numerous recent studies have been published in the literature on the development of EH systems to devise self-sufficient SHM systems. For instance, McCullagh \textit{et al}.~\cite{McCullagh2014} investigated the potential use of an electromagnetic vibration-to-electrical power generator to harvest ambient vibrations from the New Carquinez Bridge in California, US. Their results reported that the harvester generated an average power ranging between 1.6 to 5 $\upmu$W, showing promising potential for complementing other harvesting sources such as solar panels. Li and Jing~\cite{Li2019} presented the design of a piezoelectric energy harvester with a compression-to-compression force amplification mechanism for harvesting mechanical energy from highway traffic. Field tests informed that, under the passage of a mid-size vehicle at a speed of 40 km/h, the developed harvester produced an electric energy of 20.29 J and maximum open-circuit voltage of 484 V, sufficient to charge a DC battery. Another noteworthy contribution was made by Khalili and co-authors~\cite{Khalili2022} who reported the development of a self-sustained WIM system powered by vehicular traffic forces. The developed system utilized two sets of piezoelectric stacks for energy harvesting and traffic monitoring. Each stack comprised four stacks connected in parallel, and formed in turn by six PZT elements of alternating polarity connected electrically in parallel. The presented results reported a power generation of 244 mW (64\% efficiency) under sinusoidal loads of 13 kN amplitude and 80 Hz frequency (equivalent to 100 km/h). Despite the considerable number of applications in the literature, EH remains an active and growing area of research~\cite{Sharma2022,Sezer2021}. Among others, open challenges include the development of new harvesting devices with higher output power, low-power electronics, more efficient energy collection and storage systems, and non-linear piezoelectric devices apt for broadband harvesting.

\section{Background and objective}

This paper presents a novel proof-of-concept self-sensing pavement for self-sustained WIM applications through vibration-based energy harvesting of traffic loads. This work follows the previous experience by some of the authors on the development of piezoresistive composite asphalt, \rd{which spans from its material characterization to its real-life field application}. In particular, the development and characterization of a new self-sensing asphalt made of non-bituminous thermoplastic binder doped with natural aggregates and carbon microfibers were reported in reference~\cite{Birgin2021}. Later, the potential of this composite asphalt as a low-cost sensor for WIM was investigated in reference~\cite{Birgin2022}. The analyses covered the identification of 82 factory trucks over two months, achieving errors below 20\% and compliant with most international standards.

\rd{The previous research work also evidenced the high compatibility of the developed smart pavements as WIM sensors with custom-made low consumption electronics. In this light, customized electronic devices for resistivity measurements can operate at a minimum energy consumption owing to the high electrical resistance of the developed composite pavement material. Specifically, the composite pavement sensors possess resistance values around 10-100 k$\Omega$, resulting in electrical current demand below the milliampere level for operational voltage levels around 3-5 V~\cite{Birgin2021}. For comparison, note that traditional load cells operate at resistance values around 100-300 $\Omega$, having electrical current demand over 10 milliamperes if operated with the aforementioned voltage levels. Motivated by the previously observed low power consumption by these smart composites, the present work explores the potential of vibration-based EH to provide sufficient power to develop self-sustained WIM systems. The present study} has been conceptualized at the prototyping level and tested under laboratory conditions, \rd{with special emphasis on the fundamental aspects for the field application. These include (i) the repeatability and strain sensitivity of the smart pavement, (ii) balance between the energy consumption of the smart pavement and the EH production, (iii) potential of the EH system to be tuned to the specific on-site vibration conditions, and (iv) robustness of the custom-made electronics}. The investigated energy \rd{harvesting unit} consists of two piezoelectric bimorph cantilever beams connected in series. The dynamic response of the harvester beams is characterized through an electromechanical finite element model (FEM), later verified through vibration testing. \rd{A power management circuit ``full-bridge rectifier'' consisting of diodes} is tailored to convert the AC outputs of the harvesters and convert it into a usable DC voltage. The EH system is integrated with a bespoke DAQ and sensing system. The electrical output of the self-sensing pavement is operated within a dedicated circuit connected to a battery and a shunt resistor in series. The performance of the developed system is tested under laboratory conditions with harmonic and traffic loading conditions. The presented results suggest the feasibility of developing self-sustained smart pavements for long-term traffic monitoring and WIM. \rd{Overall, the main original contributions of this work comprise (i) the development of a new self-powered data acquisition system, (ii) a novel approach for the fabrication and electromechanical testing of the piezoresistive composite pavement, and (iii) laboratory feasibility analysis of the EH unit to conduct traffic load identification through electrical resistivity measurements of the smart pavement.}

The remainder of this work is organized as follows. Section~\ref{SecT00} presents the proposed concept of self-powered smart pavements. Section~\ref{SecT1} overviews the fundamentals of strain self-sensing pavements and piezoelectric bimorph cantilever beams. Section~\ref{SecT2} describes the designed DAQ system and instrumentation. Section~\ref{SecT3} presents the numerical results and discussion and, finally, Section~\ref{SecT4} concludes the paper.

\section{Self-sustained self-sensing pavements: Technology Concept}\label{SecT00}

\begin{figure}[H]
    \centering
    \includegraphics[scale = 0.7]{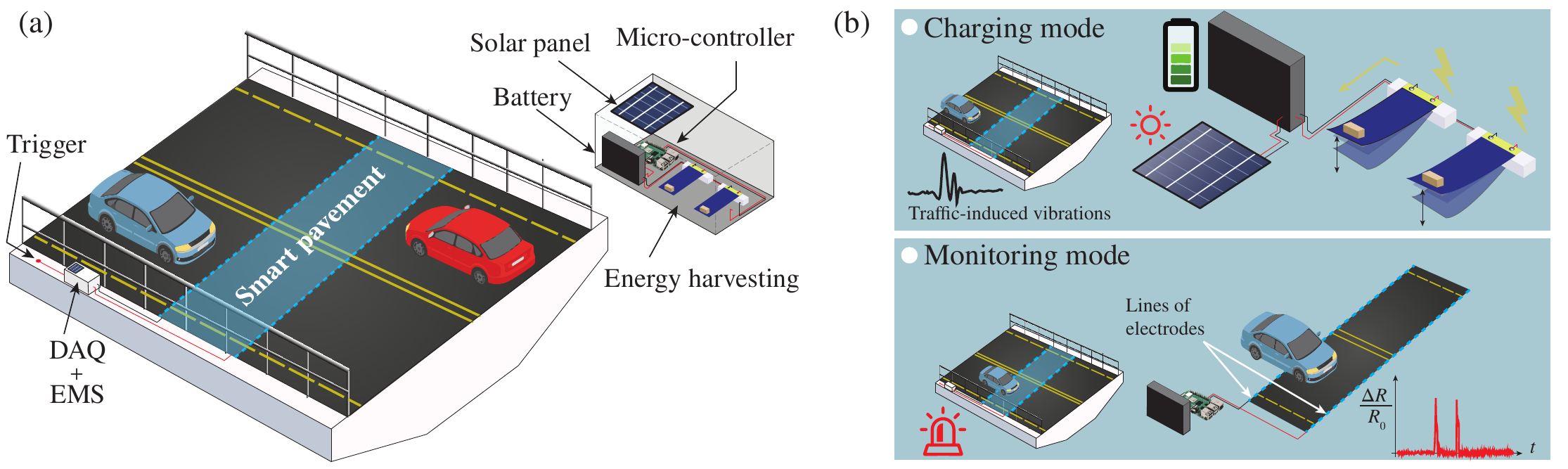}
    \caption{Technology concept of self-sensing pavements for long-term self-sustained traffic and WIM monitoring (a). Main functioning modes (b).}
    \label{Concept}
\end{figure}

The proposed technology of self-powered smart composite pavements for traffic and WIM monitoring is sketched in Fig.~\ref{Concept} (a). The system is constituted by three main sub-systems: (i) DAQ and energy management system (EMS), (ii) triggering system, and (iii) self-sensing pavement. The DAQ/EMS system would be typically mounted in a protected roadside cabinet. The DAQ sub-system is defined by a micro-controller and associated circuitry for data acquisition and transfer, while the EMS sub-system is composed of a rechargeable battery and EH devices. In general, solar energy harvesters may produce sufficient power on sunny days for regular sensor data collection and transmission. Nevertheless, their power production is limited or non-existent during night-times, cloudy periods, or when excessive dust accumulates on the panels. Thereby this work explores the use of vibration-based energy harvesters to complement or substitute solar panels. The triggering system consists of a vehicle detection sensor that activates the DAQ system to begin data acquisition. \rd{To this aim, a broad variety of sensors can be implemented such as ultrasonic sensors, magnetometers, MEMS accelerometers, infra-red sensors, or induction-loops, to mention a few. These sensors can be tuned to activate the DAQ only when a specific threshold is surpassed (corresponding to certain vehicles or trucks), so increasing the energy efficiency of the system}. The traffic load sensing is performed by the smart pavement, which represents a section cut of self-sensing composite material with a continuous connection with the road pavement. The smart pavement possesses piezoresistive properties, in such a way that its strain condition under the passage of a vehicle can be inferred from electrical resistivity measurements conducted between pairs of embedded electrodes.  

According to the conceptualized WIM set-up in Fig.~\ref{Concept}(a), two functioning modes are distinguished as illustrated in Fig.~\ref{Concept}(b), namely charging and monitoring modes. The charging mode corresponds to the charging of the battery in the EMS system by the energy harvesters. In general, the EH system may contain a solar panel and vibration-based harvesters exploiting ambient and traffic-induced vibrations. This mode will be continuously in operation except for time instants when the monitoring mode is active. The monitoring phase is activated by the trigger sensor when a vehicle approaches the monitored section. When a vehicle passes by, the pavement sensor exhibits a variation in its intrinsic electrical conductivity by virtue of its piezoresistive property. This manifests as an instantaneous variation in the electrical current crossing the pavement when subjected to a stable potential difference. Once the vehicle passage is registered, the monitoring mode is set off and the DAQ system is set back to low-power consumption ``sleeping mode'' until the next triggering event. 

\section{Theoretical background}\label{SecT1}

Firstly, the configuration and testing set-up of the investigated smart pavement, as well as a concise overview of the theoretical principles governing its strain self-sensitivity are presented in Section~\ref{SecT1a}. Afterwards, details on the electromechanical FEM of the investigated piezoelectric cantilever harvesters are reported in Section~\ref{SecT1b}.
 
\subsection{Self-diagnostic smart pavement}\label{SecT1a}

\begin{figure}[H]
    \centering
    \includegraphics[scale = 0.8]{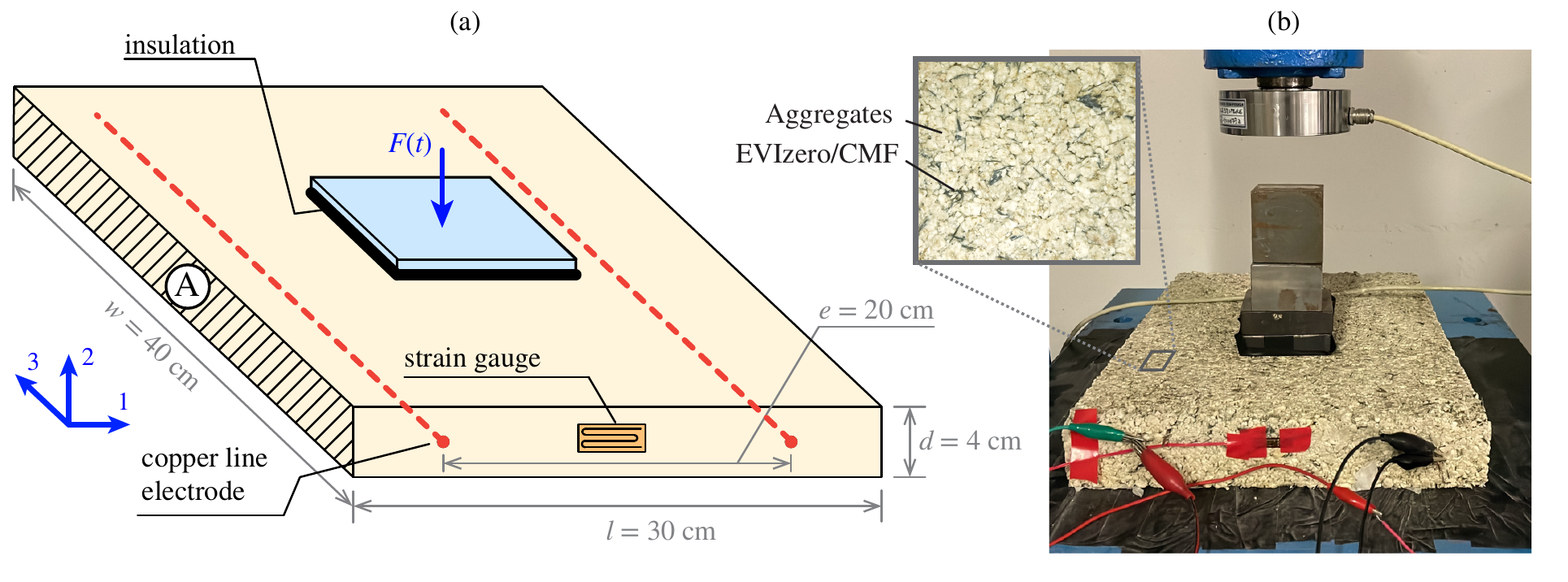}
    \caption{Investigated smart slab sample; (a) geometry and instrumentation; (b) test set-up for sensing characterization.}
    \label{slabgeo}
\end{figure}

The smart composite pavement used for traffic load sensing is made of natural aggregates (Ancona Bianco aggregates obtained from SINTEXCAL s.r.l.) held together by an admixture of a commercial binder material called EVIzero (Corecome s.r.l.~\cite{evizero}) doped with CMF (SGL Carbon~\cite{sigrafil}). The addition of CMF allows the formation of conductive networks in the microstructure of the binder, conferring electrical conductivity and piezoresistivity properties to the resulting composite asphalt. The manufacture and electromechanical characterization of the material were discussed in depth in previous research by the authors in reference~\cite{birgin2022self}. An important conclusion of that work regarded the characterization of the percolation threshold (onset of the formation of conductive networks) at filler contents proximate to 1\%. Therefore, a concentration of 1\% CMF with respect to the weight of the binder was selected to achieve maximum strain sensitivity. The designed admixture was used to fabricate a composite slab with dimensions of $40 \times 30 \times 4 \:\textrm{cm}^3$. Compression loads $F(t)$ along the 2-direction are applied through an hydraulic press onto the slab in the middle of the upper surface through a metal plate with a 10$\times$10 cm$^2$ contact area, including an insulating plastic layer between the punch and the slab to avoid electrical interferences. The sensor slab has embedded copper line electrodes with a separation of 20 cm. The slab is also instrumented with a 2-cm mono-axial strain gauge, which is located horizontally between the electrodes to measure the deformation caused by applied surface loads. The slab illustration and the characterization test set-up is depicted in Fig.~\ref{slabgeo}.

According to the notation introduced in Fig.~\ref{slabgeo}(a), the electrical resistance between the electrodes, $R$, can be formulated as~\cite{birgin2022self}: 

\begin{equation}\label{Eq_res_1}
    R = \rho \frac{e}{A},
\end{equation}

\noindent where $\rho$ is the resistivity of the composite, $e$ measures the distance between the electrodes, and $A$ denotes the cross sectional area perpendicular to the current flow. The dependency of the intrinsic electrical resistance of the slab with its strain state can be obtained by taking total derivatives of Eq.~(\ref{Eq_res_1}):

\begin{equation}\label{Eq_res_2}
\frac{\textrm{d} R}{R} = \frac{\textrm{d} \rho}{\rho} + \frac{\textrm{d} e}{e} - \frac{\textrm{d} \textrm{A}}{\textrm{A}} = \frac{\textrm{d} \rho}{\rho} + \frac{\textrm{d} e}{e} - \frac{\textrm{d} d}{d} - \frac{\textrm{d} w}{w} = \frac{\textrm{d} \rho}{\rho} + \varepsilon_1 - \varepsilon_2 - \varepsilon_3,
\end{equation}

\noindent where dimensions $l$, $d$, and $w$ are the length, thickness, and width of the slab, respectively. Given the loading configuration along the 2-direction as illustrated in Fig.~\ref{slabgeo}, the relationship between the strain components along the three directions can be readily described assuming linear isotropic elasticity as $\varepsilon_3 = \varepsilon_1  = -\nu\varepsilon_2$, with $\nu$ being the Poisson's ratio of the composite material. Then, Eq.~(\ref{Eq_res_2}) can be rewritten in terms of the strain component along the 1-direction ($\varepsilon_1$) as:

\begin{equation}
    \frac{\textrm{d} R}{R} = \frac{\textrm{d} \rho}{\rho} + \varepsilon_1 + \frac{\varepsilon_1}{\nu} - \varepsilon_1 = \frac{\textrm{d} \rho}{\rho}  - \frac{\varepsilon_1}{\nu}. \label{simplified}
\end{equation}

The strain sensitivity of the smart pavement can be quantified in terms of the GF $\lambda$ relating the relative variation of the electrical resistance $\textrm{d} R/R$ with respect to the applied strain $\varepsilon_1$ as:

\begin{equation}
\lambda = \frac{\frac{\textrm{d} R}{R}}{\varepsilon_1} = \frac{\frac{\textrm{d} \rho}{\rho}}{\varepsilon_1} +\frac{1}{\nu}.\label{piezores}
\end{equation}

Note that the GF $\lambda$ in Eq.~(\ref{piezores}) accounts for two contributions: (i) the fractional change in resistivity, $\textrm{d}\rho/\rho$, that is the piezoresistivity of the material determined by the mixture design; and (ii) the variation caused by the body deformation. It is important to remark that a GF of 3133 was experimentally obtained in reference~\cite{birgin2022self} for cylindrical samples of the same composite material with dimensions of 10 cm diameter and 6 cm height.

\subsection{\rd{Numerical model of piezoceramic harvesters}}\label{SecT1b}

\begin{figure}[H]
    \centering
    \includegraphics[scale = 1.0]{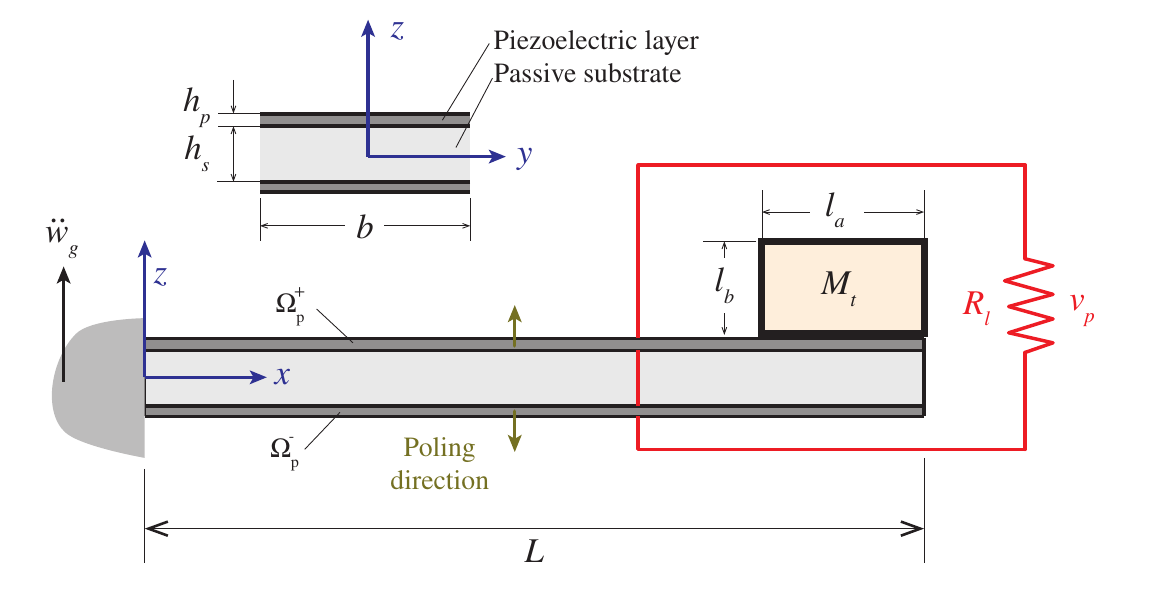}
    \caption{Bimorph piezoelectric cantilever beam model with series configuration.}
    \label{Beam_model}
\end{figure}

In this work, the investigated energy harvesters correspond to symmetric bimorph cantilever beams. As sketched in Fig.~\ref{Beam_model}, the composite cross-section is defined by two identical active piezoelectric layers of thickness $h_p$ separated by a dielectric passive substructure of thickness $h_s$. Let us denote the length and width of the beam with $L$ and $b$, respectively. The layers are assumed to be perfectly bonded with no relative sliding at the interfaces, and the ensemble is subjected to a base motion input represented by a translational displacement $w_g(t)$. The longitudinal and transverse axes are denoted by $x$ and $z$, respectively, so the neutral surface in the undeformed configuration is coincident with the $xy$-plane. Perfectly conductive electrodes are assumed to cover completely the top and bottom faces of the piezoceramic layers so that a single electric potential difference \rd{$v_p$} is defined across the active layers \rd{(i.e. the voltage drop between the two terminals of the electrodes embedded in the pavement)}. The circuit is completed with an external resistive electrical load $R_l$. Typically, proof masses are added to the harvester to decrease its fundamental frequency towards the most frequently excited broadband depending on the specific application. The incorporation of a tip mass into the electromechanical model of the harvester is later discussed in Section~\ref{SecT3a}.

Hypotheses of small deformations and linear elastic material behaviour are assumed, thus material, geometric and dissipative nonlinearities are disregarded~\cite{Stanton2010a,Stanton2010}. The substructure layer is isotropic and the piezoceramic layers are transversely isotropic as they are poled in the thickness direction. In this work, the active layers are assumed to be connected in series to produce larger voltage outputs. In this configuration, the active layers are poled oppositely in the thickness direction ($z$-direction) as illustrated in Fig.~\ref{Beam_model}. Under the assumption of linear piezoelectricity, the linear constitutive relations read~\cite{Smits1991}:

\begin{equation}\label{FEM_beam3}
\begin{split}
\bm{\sigma} &= \textbf{C} \, \bm{\varepsilon}+\textbf{e} \, \textbf{E},\\
\textbf{D} &= -\textbf{e}\bm{\varepsilon}+\bm{\epsilon}\textbf{E},\\
\end{split}
\end{equation}

\noindent where terms $\textbf{C}$, $\bm{\varepsilon}$, $\bm{\sigma}$, $\textbf{e}$ and $\bm{\epsilon}$ represent the stiffness, strain, stress, piezoelectric, and dielectric tensors, respectively. Terms $\textbf{D}$ and $\textbf{E}=-\nabla \phi$ stand for the vectors of electric displacement components and the electric field components, respectively, and $\phi$ is the electric potential. \rd{The bimorph beam is discretized in this work using 2-nodes Euler-Bernoulli beam elements as detailed in \ref{appendix:FEM}. The electromechanical governing equation of motion of the piezoelectric cantilever beam results in:}

\begin{equation}\label{FEM_beam19B}
\begin{split}
\textbf{M}\ddot{\textbf{d}}+\textbf{C}\dot{\textbf{d}}+\textbf{K}\textbf{d}-\tilde{\bm{\Theta}} \, v_p &= \textbf{F},\\
\left(\tilde{\bm{\Theta}}\right)^\textrm{T}\dot{\textbf{d}}+C_p \, \dot{v}_p+\frac{v_p}{R_l}=0.\\
\end{split}
\end{equation}

\noindent \rd{where $\textbf{M} \in \mathbb{R}^{n_m \times n_m}$ is the global mass matrix,  $\textbf{K} \in \mathbb{R}^{n_m \times n_m}$ is the global stiffness matrix , $\tilde{\bm{\Theta}} = \textbf{1}_{n_e \times 1}\bm{\Theta} \in \mathbb{R}^{n_m \times 1}$ is the electromechanical coupling matrix, $C_p$ is the capacitance of the beam, $\textbf{F}$ is the global vector of mechanical forces, $\textbf{d}\in \mathbb{R}^{n_m \times 1}$ is the global vector of mechanical displacements, and $R_l$ is the selected resistance value of the shunt resistor}.

\section{DAQ/EMS system and instrumentation of smart pavements for long-term WIM monitoring}\label{SecT2}

The developed EMS system comprises four rechargeable Ni-MH AA batteries (2000 mAh, 1.2 V) and an EH system made of two bimorph piezoelectric cantilever beams connected in series. Specifically, pre-mounted and wired bending piezoelectric generators (model Q220-A4BR-2513YB) from Piezo Systems Inc. are used in this research. At the high-level design of this system, the harvester, the battery, and the monitoring sub-system are functioning simultaneously. The system includes two circuits in parallel, namely the energy harvesting circuit and the sensing circuit, both connected to the battery. 

\begin{figure}[H]
    \centering
    \includegraphics[width = 0.5\textwidth]{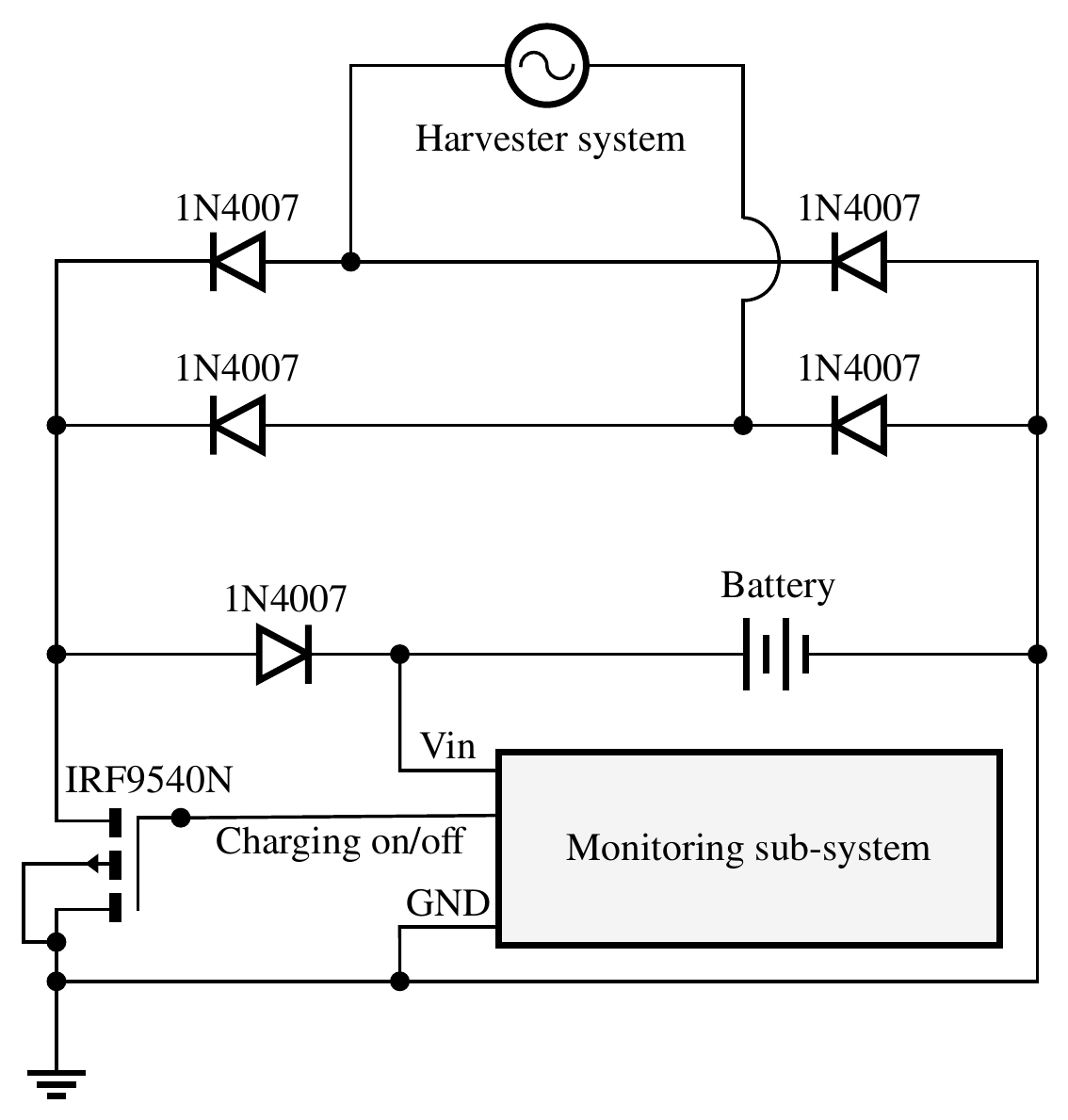}
    \caption{Electronic circuit block diagram of EH system.}
    \label{harvscheme}
\end{figure}

The EH circuit is sketched in Fig.~\ref{harvscheme}. The circuit involves a bridge rectifier and a digitally controllable metal–oxide–semiconductor field-effect (MOSFET) transistor, which controls the transitions between charging and monitoring modes. During the charging mode, the EH system generates an alternating current (AC) that is converted to direct current (DC) via a full bridge rectifier consisting of four 1N4007 diodes, and the generated DC flows to the battery via a 1N4007 diode. Nevertheless, during the sensing mode, the monitoring sub-system requires a stable voltage from the battery without being influenced by the outputs of the energy harvesters. To achieve this, the P-channel MOSFET IRF9540N is connected between the rectified output of the harvesters and the ground. The MOSFET switch is controlled by the output of the monitoring sub-system. In this way, the switch remains open to activate the charging mode, while it is held closed during the sensing period and the battery is kept separated from the ground by the diode in between. Overall, the developed EH circuit provides input voltage (Vin) and ground (GND) to the monitoring sub-system and receives 5 V logic signals (trigger) to control the functioning mode (battery charging on/off). \rd{The 5V logic voltage level is specific to the ATMega328PU micro-controller and related circuitry used in the DAQ, as explained below.}

\begin{figure}[H]
    \centering
    \includegraphics[width = 1\textwidth]{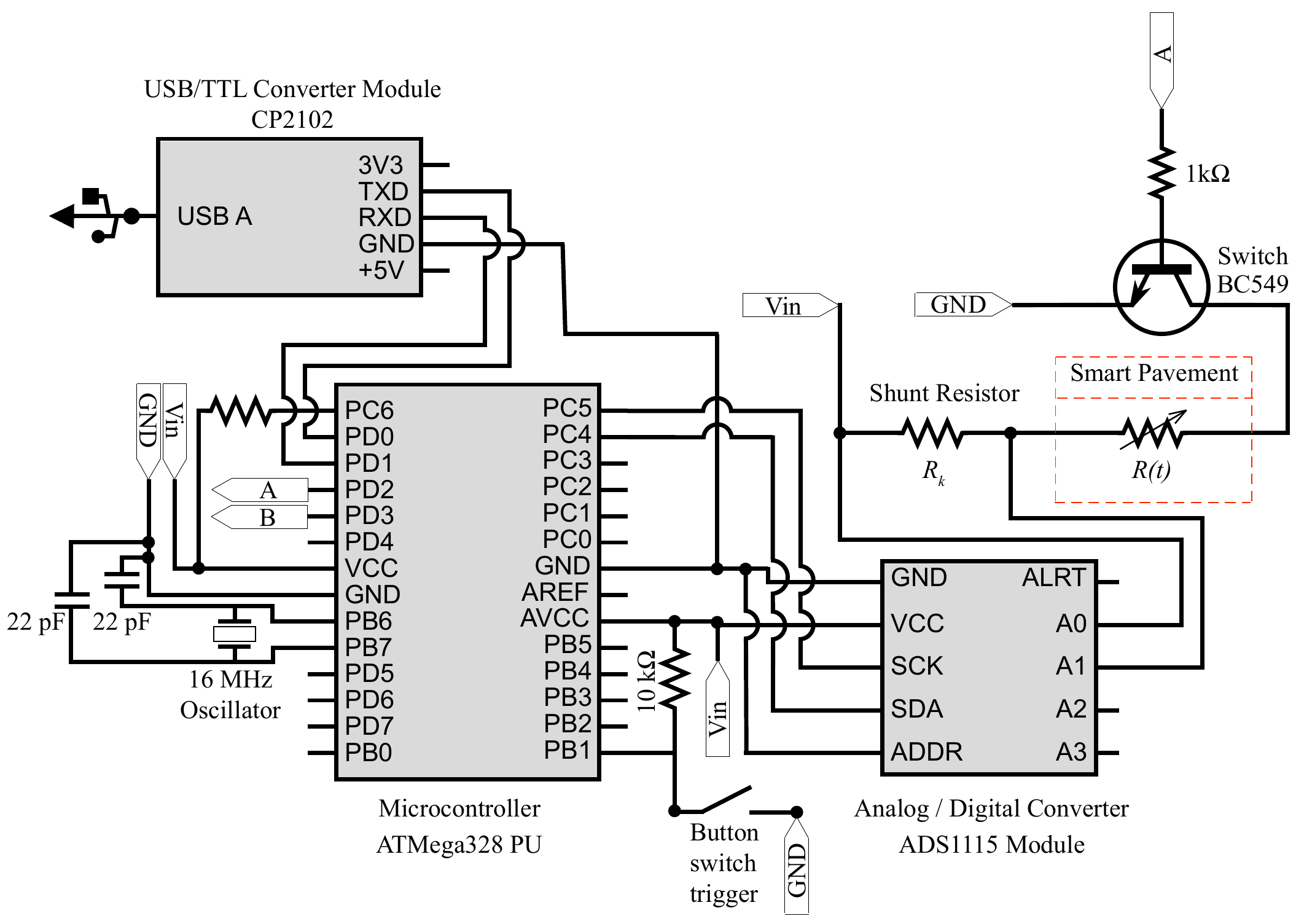}
    \caption{\rd{Electronic} circuit block diagram of DAQ system.}
    \label{daqscheme}
\end{figure}

The schematics of the developed monitoring sub-system are presented in Fig.~\ref{daqscheme}. The traffic-induced deformation of the smart pavement is assessed through a two-probe DC resistivity measurement scheme. To do so, the smart pavement is connected in series with a shunt resistor forming the sensing circuit, together with a low-side transistor switch BC549 acting as a digital on/off control. When the monitoring mode is activated and a stable DC flows between the electrodes of the pavement, the DAQ system records the instantaneous voltage drop $V_k(t)$ between the terminals of the shunt resistor and transmits the measurements via universal asynchronous receiver transmitter (UART) communication. It is important to remark that the proposed composite material exhibits no polarization effects when subjected to DC potential differences as previously reported in reference~\cite{birgin2022self}. Therefore, no significant drifts in the electrical resistance of the composite are observed when subjected to DC, which considerably simplifies the assessment compared to AC schemes. The resistance time history of the pavement, $R(t)$, can be readily computed by the direct application of the Ohm's law as:

\begin{equation}\label{Ohm}
R(t) = R_k \frac{V_p(t)}{V_k(t)},
\end{equation}

\noindent where $V_p(t)$ is the voltage drop between the two terminals of the electrodes embedded in the pavement, and $R_k$ is the selected resistance value of the shunt resistor.

To maximize the energetic efficiency of the system, the DAQ system stays in stand-by (sleeping) mode during the charging phase. Once the trigger activates the monitoring mode, the system reactivates and acquires data at the sampling rate of 10 Hz for 10 seconds, which suffices to monitor the complete passage of vehicles travelling at moderate speeds over the sensing pavement. \rd{In the results presented hereafter in Section~\ref{SecT3}, the trigger was manually operated. Note however that it can be easily managed by external logic signals with very low energy demand. For instance, it is possible to use a low power consumption MEMS accelerometer activating the input pin when the acceleration overpasses a specific threshold (properly tuned to identify heavy trucks). Nonetheless, since this aspect does not compromise the feasibility of this technology, this detail in the design is left for future development stages}. The DAQ system designed in this study consists of an ATMEGA328PU micro-controller, ADS1115 analog-to-digital converter from Texas Instruments, and a CP2102 USB/TTL Converter module. The schematic includes the basic connections for ATMEGA328PU to operate at 16 MHz adapted from the data-sheet of the micro-controller. The outputs indicated by A and B in Fig.~\ref{daqscheme} denote the digital outputs that control the current of the sensing circuit and charging, respectively. Finally, the serial outputs of the DAQ are collected via a USB connection through a dedicated script created in Python language. 

\begin{figure}[H]
    \centering
    \includegraphics[scale = 1.2]{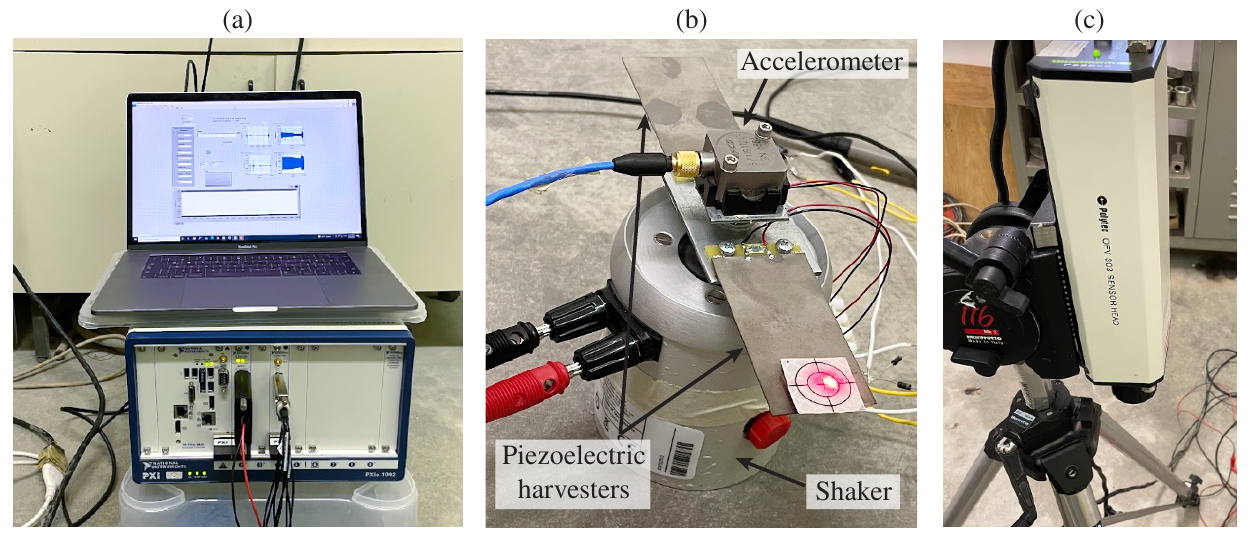}
    \caption{Experimental set-up for dynamic characterization of energy harvesters: (a) computer and data logger, (b) vibration-based energy harvesters connected to shaker, (c) laser Doppler vibrometer.}
    \label{Experiment_nomass}
\end{figure}

During the experimental campaign, a National Instruments PXIe has been used for monitoring and evaluation purposes. The voltage time histories of the battery have been recorded by NI PXIe 4032 analog-to-digital converter ($\pm$10 V, resolution 50 $\upmu$V), and the time series of electrical current on the battery power-line have been monitored by NI PXIe 4071 digital multimeter ($\pm$20 mA, resolution 10 nA) (Fig.~\ref{Experiment_nomass} (a)). Readings from the strain guage have been acquired through NI-PXIe 4330 bridge input module. The energy harvesters were screwed to a rigid clamp, screwed in its centre to an electrodynamic permanent magnet shaker (Br{\"u}el \& Kjaer, model V201, 5 Hz-13 kHz, peak force 26.7 N) actuated by using NI PXIe 4138 voltage source for simulating traffic-induced vibrations (Fig.~\ref{Experiment_nomass} (b)). Vertical accelerations at the base of the harvesters have been recorded by a small MEMS accelerometer (PCB 3711B112G, $\pm$2 g, 1 V/g) attached to the clamp of the cantilever with a small magnet and connected to the NI Sound and Vibration module PXIe 4492. The tip velocity of the harvesters in the transverse direction has been recorded by a laser Doppler vibrometer (Polytec OFV-3001/OFV-303) (Fig.~\ref{Experiment_nomass} (c)) by attaching a small piece of reflector tape at the tip, and the measurements were recorded with the NI vibration module NI PXIe 4492. The load during the electromechanical characterization test has been applied through a manually operated hydraulic press (Controls 50-C7600) with load capacity of 15 kN. The load time history has been measured through LAUMAS load cell with a 10 kN maximum reading capacity read by NI-PXIe 4330. All National Instruments devices have been controlled by tailor-made LABVIEW software codes.

\section{Numerical Results and Discussion}\label{SecT3}

The experimental campaign has been organised in three stages. Section~\ref{SecT3a} presents the results of the characterization testing of the smart pavement and the bimorph cantilevered beams used in the EH system. Section~\ref{SecT3b} evaluates the power generation capability of the developed EH system and, finally, Section~\ref{SecT3c} assesses the performance of the developed WIM system to identify harmonic and traffic loads. The electromechanical properties of the piezoelectric cantilevered harvesters and the material properties of the smart composite pavement are collected in Table~\ref{prop_harve}. 

As anticipated above, the smart composite pavement is made of natural aggregates, EVIzero and CMF. EVIzero is a neutral-coloured eco-friendly binder based on polyolefin, made with polymers and industrial by-products. The aggregates include 0-4 mm sized fine aggregates and 4-8 mm sized coarse aggregates complying with minimum/maximum design gradation specifications for roadway pavements~\cite{Corecom_car}. The CMFs are cut carbon fibers with a single fiber length of 6 mm and a diameter of 7 $\upmu$m (aspect ratio $\approx$860). These fillers are highly conductive, with single filament resistivity of 15 $\upmu \Omega$m, in such a way that low filler contents may yield increases in the binder several orders of magnitude higher than the pristine material. The production process of the pavement material was similar to the one of bitumen road asphalt, including (i) preparation and heating of the material, (ii) mechanical mixing, and (iii) compaction. The aggregates and EVIzero were first heated in the oven under 180 $^\circ$C for 3 hours and 1 hour, respectively. Then, the admixture was blended through mechanical mixing using a laboratory mixer. The admixture was then poured in a mould in two layers of equal thickness, deploying the line electrodes in between. Finally, the material was compacted using a roller compactor (Controls 77-PV41A02) until achieving an uniform thickness of the slab of 4 cm, which approximately corresponds to a mass density of 2500 g/cm$^3$. For further details on the manufacturing process and material characterization, readers may refer to reference~\cite{birgin2022self}.

\begin{table}[H]										
\setlength{\tabcolsep}{3pt} 
\newcommand\Tstrut{\rule{0pt}{0,3cm}}         
\newcommand\Bstrut{\rule[-0.15cm]{0pt}{0pt}}   
\footnotesize										
\caption{Electromechanical properties of bimorph cantilevered harvesters (\rd{reference}~\cite{Jaffe1958} and technical data-sheet of Q220-A4BR-2513YB) and material properties of smart composite pavement~\cite{evizero,sigrafil,birgin2022self}.}  										
\vspace{0.1cm}										
\centering										
\begin{tabular}{lrllrr}										
\hline										
\multicolumn{2}{l}{Cantilevered bimorh energy harvesters}			& 	& 	\multicolumn{2}{l}{Smart pavement}					\Tstrut\Bstrut\\
\cline{1-2}\cline{4-6}										
\multicolumn{2}{l}{Geometry} 			& 	& 	Material characteristics	& 	EVIzero 	& 	CMF	\Tstrut\Bstrut\\
\cline{1-2}\cline{4-6}										
Length $L$ [mm] 	& 	57.7	& 	& 	Conductivity [S/m]	&  - & 	6.67E+04	\Tstrut\\
Width $b$ [mm] 	& 	31.8	& 	& 	Density [g/cm$^3$]	& 	0.85	& 	1.8	\\
Thickness of the substrate $h_s$ [mm] 	& 	0.17	& 	& 	Mixing temperature [$^\circ$C]	& 	 160–170	&  - \\
Thickness of the piezoceramic layers $h_p$ [mm] 	& 	0.19	& 	& 	Dynamic viscosity at 160 $^\circ$C [mPa$\cdot$s] 	& 	700	&  - \Bstrut\\
\cline{1-2}\cline{4-6}										
\multicolumn{2}{l}{Constitutive properties of the piezoceramic layers} 			& 	& 	\multicolumn{2}{l}{Mix design}			& 		\Tstrut\Bstrut\\
\cline{1-2}\cline{4-6}										
Young's modulus $c_{11}$ [GPa] 	& 	66	& 	& 	Fine aggregates [wt.\%]	& 	\multicolumn{2}{r}{49.15}			\Tstrut\\
Mass density $\rho_p$ [g/cm$^3$]	 & 	8	& 	& 	Coarse aggregates [wt.\%]	& 	\multicolumn{2}{r}{44.91}			\\
Piezoelectric coefficient $e_{31}$ [C/m$^2$] 	&	-5.4	& 	& 	EVIzero [wt.\%]	& 	\multicolumn{2}{r}{5.89}			\\
Dielectric constant $\epsilon_{33}$ [nF/m] 	&	7.96	& 	& 	CMF [wt.\%]	& 	\multicolumn{2}{r}{0.06}			\Bstrut\\
\cline{1-2}										
\multicolumn{2}{l}{Constitutive properties of the substrate} 			& 	& 		& 		& 		\Tstrut\Bstrut\\
\cline{1-2}										
Young's modulus $Y_s$ [GPa]	 & 	100	& 	& 		& 		& 		\Tstrut\\
Mass density $\rho_s$ [g/cm$^3$] 	& 	8.4	& 	& 		& 		& 		\Bstrut\\
\hline 										
\end{tabular}										
\label{prop_harve}										
\end{table}    	

\subsection{Characterization testing of smart pavement and EH system}\label{SecT3a}
								
\subsubsection{Experimental characterization of smart pavement}

\begin{figure}[H]
\centering
  \includegraphics[scale = 1.00]{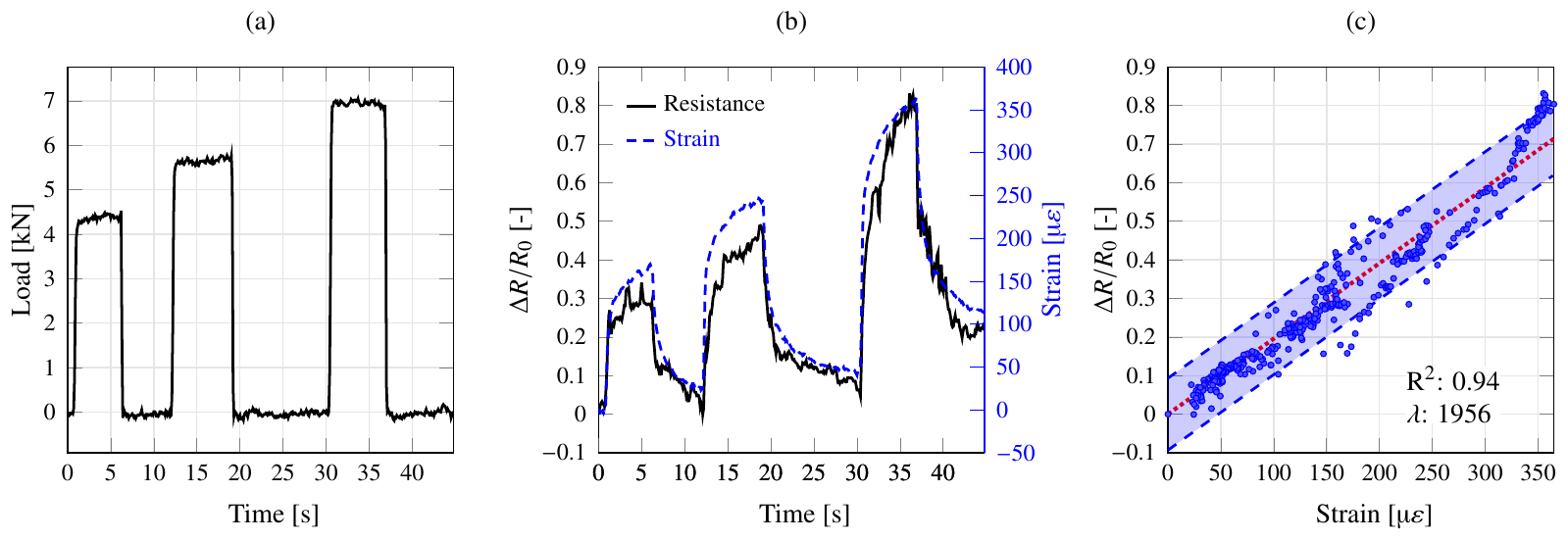}
  \caption{Electromechanical test results of smart pavement slab; (a) load time history; (b) relative variation of electrical resistance and measured strain time histories; (c) correlation analysis between relative variation of electrical resistance and mechanical strain (shaded areas represent 95\% confidence levels).}
  \label{platechar}
\end{figure}

The self-sensing property of the smart pavement slab is assessed first. To do so, the composite slab was subjected to a series of step compression loads of 4.4, 6.7, and 7 kN. The load steps were hold during $\approx$5 s to inspect the stability of the readings (Fig.~\ref{platechar}(a)). The electrical part of the tests consisted of resistance measurements of the slab using the embedded electrodes. The sensing system operated at 5 V DC supplied by the battery, and a shunt resistor of 10 k$\Omega$ was used to conduct the resistivity measurements. The obtained time series of measured strain and relative variation of the electrical resistance are reported in Fig.~\ref{platechar}(b). It is firstly observed that, in agreement with previous reseach by the authors in reference~\cite{birgin2022self}, no drifts were found as a result of polarization. This confirms the suitability of using DC resistivity measurements to conduct sensing applications. It is also evident in this figure the existence of a clear correlation between strain and resistance in good agreement with the piezoresistivity theory previously introduced in Section~\ref{SecT1a}. It is noted in both the resistance and strain signals that the amplitudes are correlated with the magnitude of the load. A closer inspection reveals the existence of a transient behaviour as the slab switches between the unloaded and loaded states, as well as the presence of residual strains at the end of the resting stages. This residual shift observed during the period between consecutive loading events is attributed to the visco-elastic nature of the material. Overall, these results demonstrate the stability and low signal-to-noise ratios of the electrical measurements.

The correlation between the mechanical strain and the induced variation in the intrinsic electrical resistance of the pavement is further investigated in Fig.~\ref{platechar}(c). Despite the presence of some dispersion in the results due to noise in the measurements (specially in the strain time series, see Fig.~\ref{platechar}(b)), it is clear in this figure that both magnitudes are highly correlated. Indeed, the linear fitting model in Fig.~\ref{platechar}(c) reports a coefficient of determination very close to 1 ($\textrm{R}^2=0.94$), which corresponds to a perfect correlation. The gauge factor of the pavement is computed as the slope of the linear fitting model, obtaining a value of $\lambda=1956$. This result is in good agreement with previous experience by the authors in reference~\cite{birgin2022self}, where gauge factors of 3133 and 1000 were reported for cylindrical samples and the slab pavement sample with similar dimensions and instrumentation, respectively, composed of the same composite material and fabricated following the identical manufacturing process. These results therefore demonstrate that (i) the developed smart pavement exhibits high strain sensitivity, and (ii) the manufacturing process assures the repeatability of the material properties. 

\subsubsection{Numerical and experimental characterization of energy harvesters}

In this section, the dynamic behaviour of the bimorph harvesters is identified in terms of frequency response functions (FRFs). In particular, two FRFs are obtained including the tip velocity to base acceleration and the voltage output to base acceleration. Chirp excitation is provided to the shaker with frequencies ranging from 10 Hz to 2 kHz in 180 seconds. Since the goal is to compare the experimental results with the linear electromechanical model previously reported in Section~\ref{SecT1b} \rd{and \ref{appendix:FEM}}, it is ensured that the base acceleration level does not overpass 0.2g so that the inherent non-linearities remain inactive (far below the $\pm1.6$g acceleration rating of the harvesters). In the results hereafter, an external resistive electrical load of value $R_l=100$ $\Omega$ has been selected.

From the FEM formulation in Section~\ref{SecT1b}, it remains to define the steady state response of the cantilevered harvester under harmonic base excitation. To do so, the mode shapes and resonant frequencies of the beam are obtained first by reducing the global system matrices into the standard eigenvalue form. In the case of piezoelectric materials, two kinds of modal properties can be obtained, namely short circuit and open circuit modal properties~\cite{Erturk2011}. The open circuit natural frequencies $\omega^{oc}$ are calculated by assuming that the resistive load is infinite (i.e.~$R_l \rightarrow \infty$), so no charge flows in the circuit. Through static condensation of the output voltage, the open circuit modal properties can be calculated as:

\begin{equation}\label{FEM_beam21}
\left(\textbf{K}^*-\omega^{oc}\textbf{M}\right)\bm{\varphi}^{oc} = 0,
\end{equation}

\noindent where $\textbf{K}^* = \textbf{K}+\tilde{\bm{\Theta}}\left(\tilde{\bm{\Theta}}\right)^\textrm{T}/C_p$. 

Instead, the short circuit natural frequencies $\omega^{sc}$ are calculated by assuming that no potential difference exists across the beam in free vibration, that is the coupling term is zero at all times. The short circuit natural modal properties can then be extracted from the following eigenvalue problem:

\begin{equation}\label{FEM_beam20}
\left(\textbf{K}-\omega^{sc}\textbf{M}\right)\bm{\varphi}^{sc} = 0.
\end{equation}

On this basis, the steady-state response of the piezoelectric bimorph under harmonic base excitation $w_g=Y_oe^{-\textrm{i} \omega t}$ ($\textrm{i}=\sqrt{-1}$) can be obtained by applying modal superposition. Considering $N_m$ modes in the simulation, the voltage FRF can be obtained as:

\begin{equation}\label{FRF_Vol}
\frac{FRF_{v}(\omega)}{\omega^2Y_o}=\frac{\sum_{r=1}^{N_m} \frac{\textrm{i}\omega \tilde{\bm{\Theta}}^\textrm{T} \bm{\varphi}^{sc}_r\left(\bm{\varphi}^{sc}_r\right)^\textrm{T}\textbf{M}\textbf{L}}{\left(\omega_r^{sc}\right)^2-\omega^2+\textrm{i}2\zeta_r\omega_r^{sc}\omega}}{\frac{1}{R_l}+\textrm{i}\omega C_p + \sum_{r=1}^{N_m} \left\{\frac{\textrm{i}\omega \tilde{\bm{\Theta}}^\textrm{T} \bm{\varphi}^{sc}_r\left(\bm{\varphi}^{sc}_r\right)^\textrm{T}\tilde{\bm{\Theta}}}{\left(\omega_r^{sc}\right)^2-\omega^2+\textrm{i}2\zeta_r\omega_r^{sc}\omega}\right\}},
\end{equation}

\noindent and the velocity FRF (relative to the base) reads:

\begin{equation}\label{FRF_vel}
\frac{FRF_{\dot{\textbf{d}}}(\omega)}{\omega^2Y_o}=\textrm{i}\omega \left[
\sum_{r=1}^{N_m} \bm{\varphi}^{sc}_r
\frac{-\left(\bm{\varphi}^{sc}_r\right)^\textrm{T} \textbf{M}\textbf{L} + \left(\bm{\varphi}^{sc}_r\right)^\textrm{T}\tilde{\bm{\Theta}}\frac{FRF_{v}(\omega)}{\omega^2Y_o}} 
{\left(\omega_r^{sc}\right)^2-\omega^2+\textrm{i}2\zeta_r \omega_r^{sc} \omega}
\right].
\end{equation}

Note that the tip velocity assessed in the experiments corresponds to the motion of the cantilever relative to the fixed reference frame of the laser. Therefore, the relative tip velocity FRF given by Eq.~(\ref{FRF_vel}) needs to be modified to express the absolute velocity relative to the fixed frame of reference ($\dot{\tilde{\textbf{u}}}$) as: 

\begin{equation}\label{FRF_vel_rel}
\dot{\tilde{\textbf{u}}} = \dot{\textbf{u}}+\dot{\textbf{u}_b} \rightarrow \frac{FRF_{\dot\tilde{{\textbf{d}}}}(\omega)}{\omega^2Y_o} = \textrm{i}\omega \frac{FRF_{\dot{\textbf{d}}}(\omega)}{\omega^2Y_o} + \frac{\textrm{i}}{\omega}.
\end{equation}

\begin{figure}[H]
\centering
  \includegraphics[scale = 1.0]{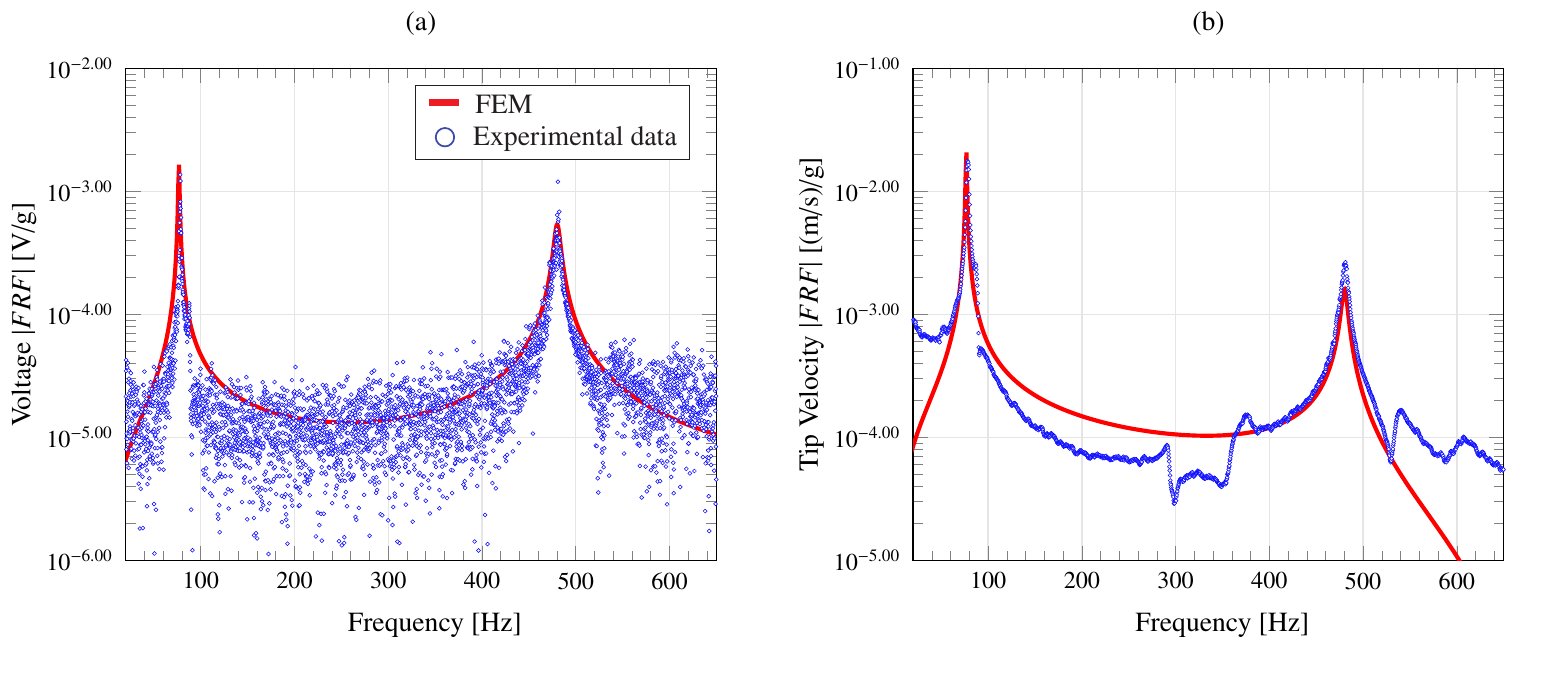}
  \caption{Comparison between experimental and numerical FRFs in terms of output voltage (a) and tip velocity (b) of a cantilevered bimorph harvester ($R_l=100$ $\Omega$).}
  \label{FRFs_notip}
\end{figure}

The experimentally measured and theoretical voltage output and tip velocity FRFs are shown in Figs.~\ref{FRFs_notip} (a) and (b), respectively. Note that, here and hereafter, the electromechanical FRFs are given in semi-log scale and normalized with respect to the gravitational acceleration for a convenient representation. The analytical model predicts the first two short-circuit resonance frequencies at 76.64 Hz and 480.25 Hz, and the open-circuit resonance frequencies at 78.73 Hz and 484.39 Hz. The experimental resonant frequencies are obtained at 77.41 Hz and 481.50 Hz by peak picking analysis of the velocity TF from Fig.~\ref{FRFs_notip} (b). Note that the selected resistive load falls in between the short- and open-circuit conditions, which explains the fact that the fundamental frequency lies in between the theoretical limits. The identification of mechanical damping is performed by matching the peaks of the experimental and analytical tip velocity FRFs in Fig.~\ref{FRFs_notip} (b), obtaining a damping ratio of $\zeta=0.69\%$. Overall, good agreements are found between the experimental and theoretical results in Fig.~\ref{FRFs_notip}, especially in the frequency broadband neighbouring the resonances. The good agreement observed for both the voltage and the tip velocity FRFs evidence the correctness of the fundamental assumptions made in the linear electromechanical model. The comparison is particularly good in terms of output voltage, despite the presence of considerable noise in the experimental data between resonances. The comparison in terms of velocity FRFs is significantly worse in this frequency range, which may be ascribed to accuracy limitations of the laser vibrometer. It is noticeable in Fig.~\ref{FRFs_notip} (b) the presence of two small resonances at 291.8 Hz and 377.7 Hz. The fundamental resonance of the shaker is about 13 kHz according to the technical specifications, therefore these low-amplitude peaks are possibly due to rotation of the clamp of the bimorph as a result of the joint flexibility at the clamp–shaker interface. 

The developed numerical model is then used to adjust the fundamental frequency of the harvester to the frequency broadband commonly excited by traffic (10-15 Hz). To do so, a prismatic mass model with cross-section $l_a \times l_b$ and perfectly attached to the tip of the bimorph cantilever is considered as sketched in Fig.~\ref{Beam_model}. Its effect is directly introduced in the global mass matrix of the beam through its mass $M_t$ and mass moment of inertia $M_t$ about the centre axis of the bimorph given by:

\begin{equation}\label{Itmass}
I_t=M_t b \left[\frac{l_a^2+l_b^2}{12}+\left(\frac{b+h_s}{2}+h_p\right)^2\right].
\end{equation}

\begin{figure}[H]
\centering
  \includegraphics[scale = 1.0]{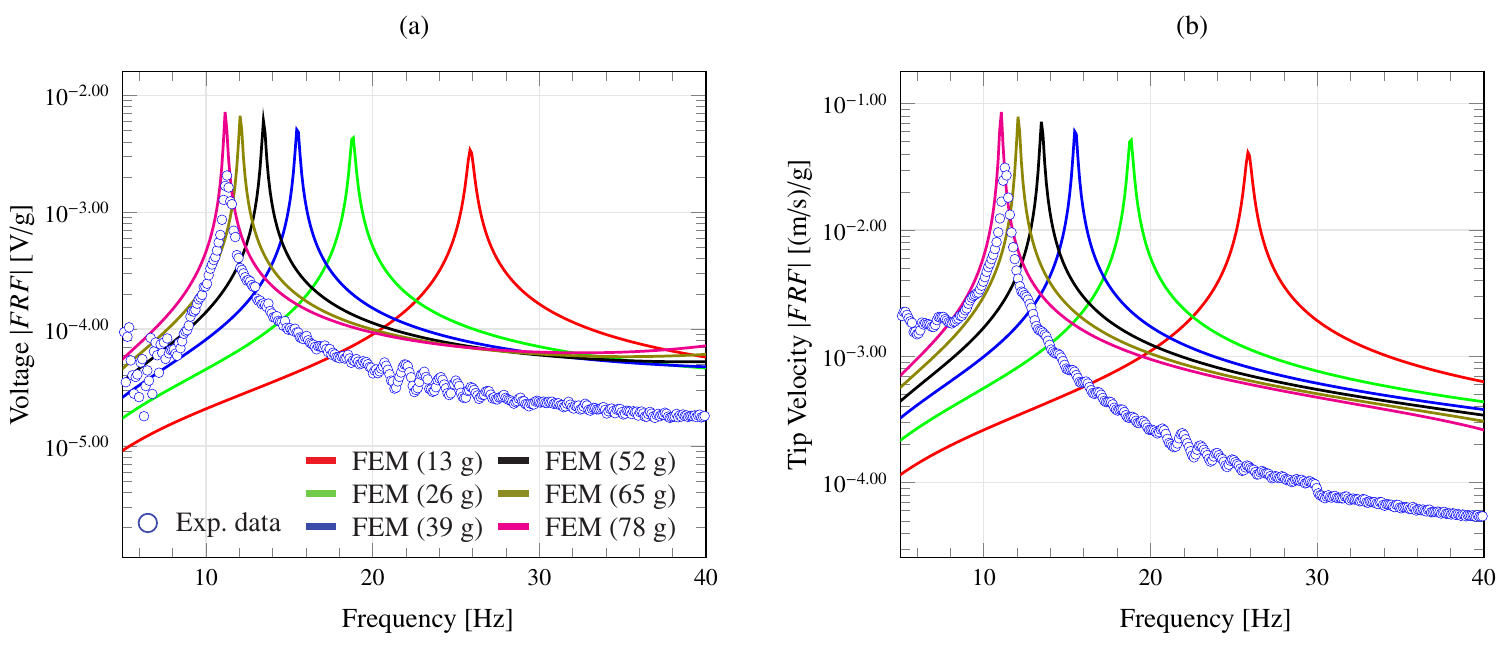}
  \caption{Comparison between experimental and numerical FRFs in terms of output voltage (a) and tip velocity (b) of a cantilevered bimorph harvester ($R_l=100$ $\Omega$). The experimental results were obtained considering a proof mass of 78 g.}
  \label{FRFs_tip}
\end{figure}

In the experiments, stacks of plane metal brackets of dimensions $l_a=14$ mm and $l_b=2$ mm and mass of 13 g \textcolor{black}{were} attached with two screws to the tip of the beam in order to have easy control over the proof mass. Figure~\ref{FRFs_tip} shows the numerical predictions of the voltage output and tip velocity FRFs considering an increasing number of metal brackets, from 1 (13 g) to 6 (78 g). The mechanical damping is kept constant from the previous analysis. It is noted how the fundamental frequency decreases from 25.84 Hz until 11.20 Hz, the latter lying within the frequency broadband of interest. This mass was considered in the experiment, obtaining a fundamental frequency of 11.3 Hz in good agreement with the numerical results. 

\subsection{Power generation of the EH system}\label{SecT3b}

\begin{figure}[H]
\centering
  \includegraphics[width = 1\textwidth]{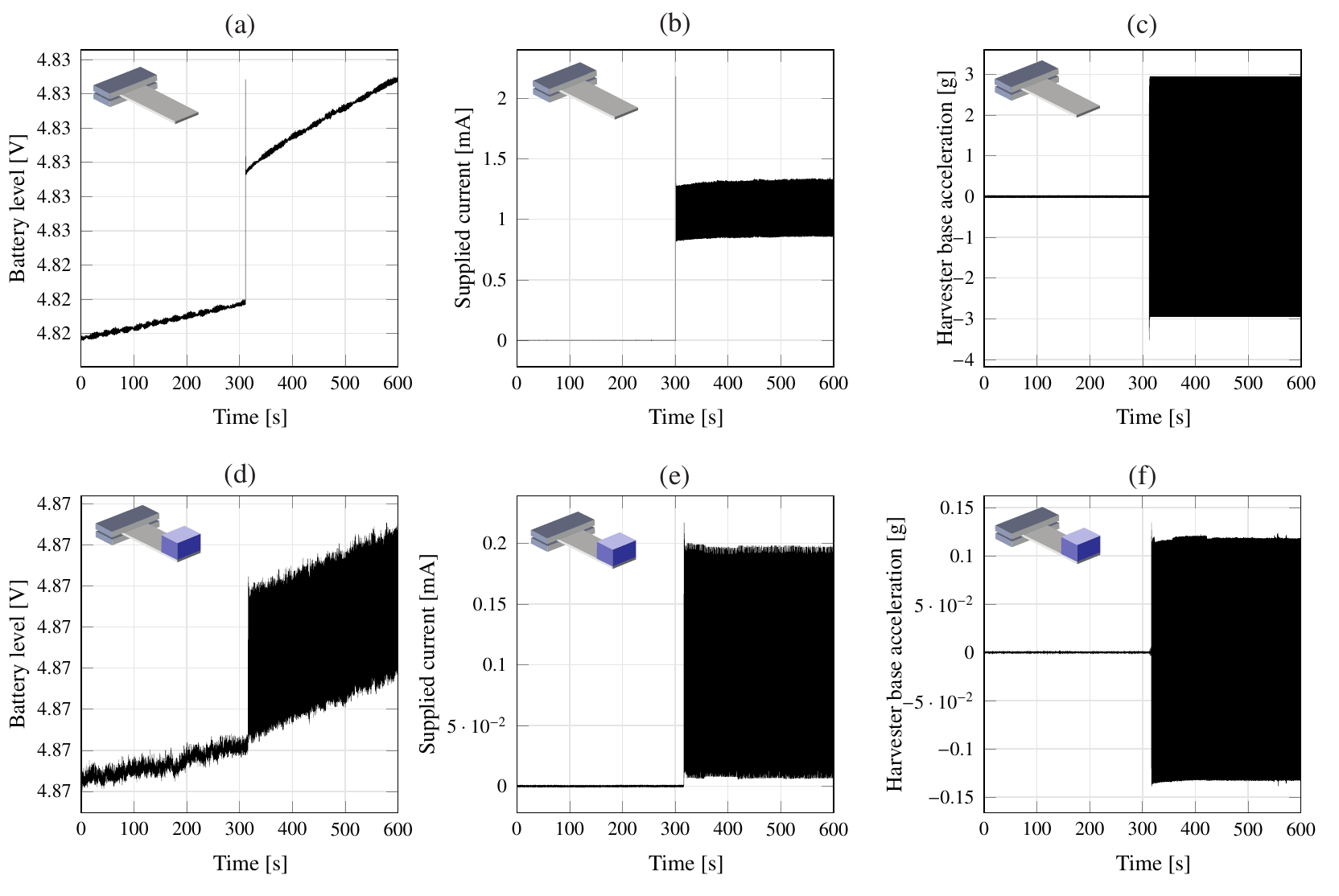}
  \caption{Battery charging performance of energy harvesters without (a,b,c) and with added proof masses (d,e,f). Voltage level of battery (a,d); Electrical current supplied by the harvesters (b,e); Vertical accelerations at the base of the harvesters (c,f).}
  \label{chargingwot}
\end{figure}

A key element of the proposed technology regards the ability of the designed EH system to charge the batteries. To test such a feature, the two cantilever beams in series are connected to the shaker and excited by harmonic base accelerations with a frequency coincident with the resonant frequency of the harvesters previously identified in Section~\ref{SecT3b}. The testing procedure consisted of 5 minutes of idle period followed by 5 minutes of harmonic loading. In the experiments, the performance of the harvesters without and with proof masses were investigated as reported in Figures~\ref{chargingwot} (a,b,c) and (d,e,f), respectively. In the case of harvesters without proof masses (Figs.~\ref{chargingwot} (a,b,c)), the vibration input was defined in the signal generator through a sinusoidal time history with frequency equal to 77.41 Hz. It is noted in Fig.~\ref{chargingwot} (a) that minor self-recovery increases are visible in the first 5 idle minutes. Once the harmonic excitation starts, the voltage state of the battery experiences a step increase with fluctuations following those in the current supplied by the harvesters (Fig.~\ref{chargingwot} (b)). In this period, the voltage time history increases with a considerably larger growth rate compared to that observed during the idle period, demonstrating the charging of the battery. In particular, the electrical power generated by the harvesters is around 1.1 mW. Nevertheless, it is observed in Fig.~\ref{chargingwot} (c) that the base acceleration required to effectively charge the battery is around 3g, which will hardly appear in real applications under traffic loading. This is considerably alleviated when incorporating the proof masses in Figs.~\ref{chargingwot} (d,e,f)). In this case, the frequency of the harmonic excitation was defined as 11.3 Hz. It is noted in Fig.~\ref{chargingwot} (c) that the charging rate of the battery appears slower than that of the configuration without proof masses in Fig.~\ref{chargingwot} (a). Nevertheless, owing the optimal dynamic amplification experienced by the harvesters when considering proof masses, it is noted in Fig.~\ref{chargingwot} (f) that only a base acceleration of 0.12g suffices to effectively charge the batteries, which can be perfectly feasible in field applications. In this case, the harvester supplied 0.11 mA to the battery which amounts to a power generation of 0.53 mW.

\begin{figure}[H]
\centering
  \includegraphics[width = 1\textwidth]{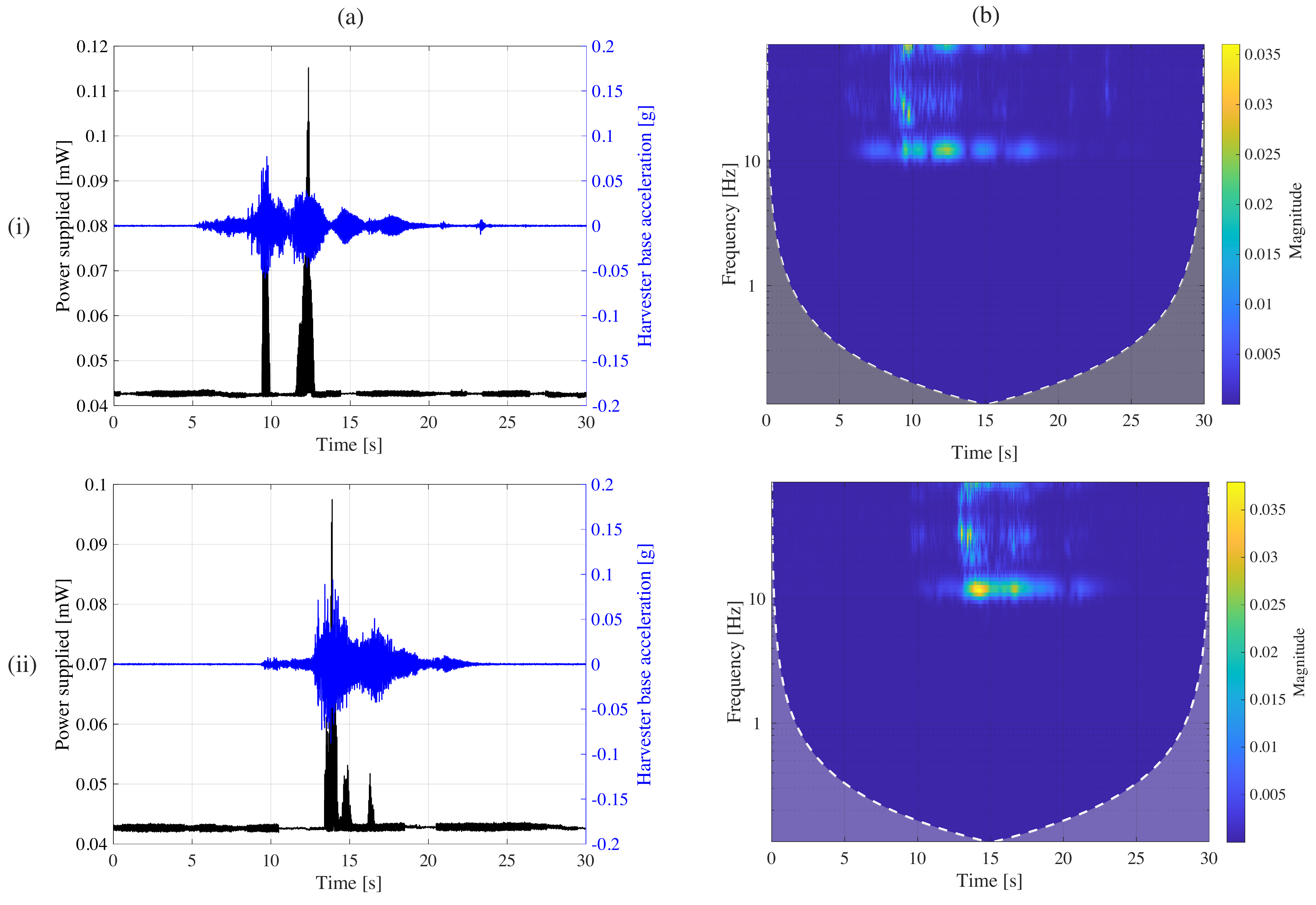}
  \caption{Power generation of the EH system under real traffic-induced vibration. \rd{The passages of two different vehicles (i-ii) are plotted together.} Input base accelerations and power generation time histories in column (a); Related time-frequency analysis of the input base accelerations in column (b).}
  \label{freqtime}
\end{figure}

In light of the previous analyses, the optimal configuration of energy harvesters with 78 g proof masses is tested under real traffic-induced excitation. To do so, ambient accelerations recorded in a real in-operation bridge are retrieved and replicated in the signal generator. In particular, acceleration time signals recorded in the Trigno V Bridge are used herein. The Trigno V Bridge is a seven span simply supported concrete-girder highway bridge ($7 \times 33.7$ m long) located in the Italian region of Abruzos between the municipalities of Isernia and San Salvo. Within the framework of research project funded by the public infrastructure manager Anas S.p.A., this bridge was instrumented on October 13$^\textrm{th}$ 2021 with a dense network of MEMS accelerometers to identify its modal properties. In particular, four asynchronous 30 min long acquisitions were acquired between 11:00 a.m. and 12:30 p.m. at a sampling rate of 200 Hz and under normal operating conditions, with wind and traffic as the main sources of excitation. In this work, one of the acceleration time signals recorded by an accelerometer located at the mid-span of the first span is selected and replicated by the shaker.  \rd{On this basis, Fig.~\ref{freqtime} (i-a) and (ii-a) report two examples the time histories of base accelerations and power generated by the harvesters under the passage of two different vehicles}. It is noted that no noticeable power is generated when the bridge experiences white noise excitations. Such vibrations are dominated by a combination of the modal signatures of the bridge, whose frequencies are far from the fundamental frequency of the harvesters. Indeed, the fundamental frequency of the first span of the Trigno V Bridge was determined at 3.8 Hz, which produces no significant dynamic amplification in the harvesters. Nonetheless, it can be clearly observed in Fig.~\ref{freqtime} \rd{(i-a) and (ii-a)} that the transient vibrations induced by the passing vehicles do effectively activate the harvesters attaining a maximum power generation of 0.11 mW. \rd{In particular, the peaks can be observed in the generated power, which conceivably correspond to the axles of the vehicles.} The time-frequency analysis of the acceleration time histories in Fig.~\ref{freqtime} \rd{(i-b) and (ii-b)} confirm the concentration of energy in the frequency broadband surrounding the fundamental frequency of the harvesters (11.3 Hz). 

\subsection{Operational evaluation of automated WIM system}\label{SecT3c}

\begin{figure}[H]
\centering
  \includegraphics[scale = 1.0]{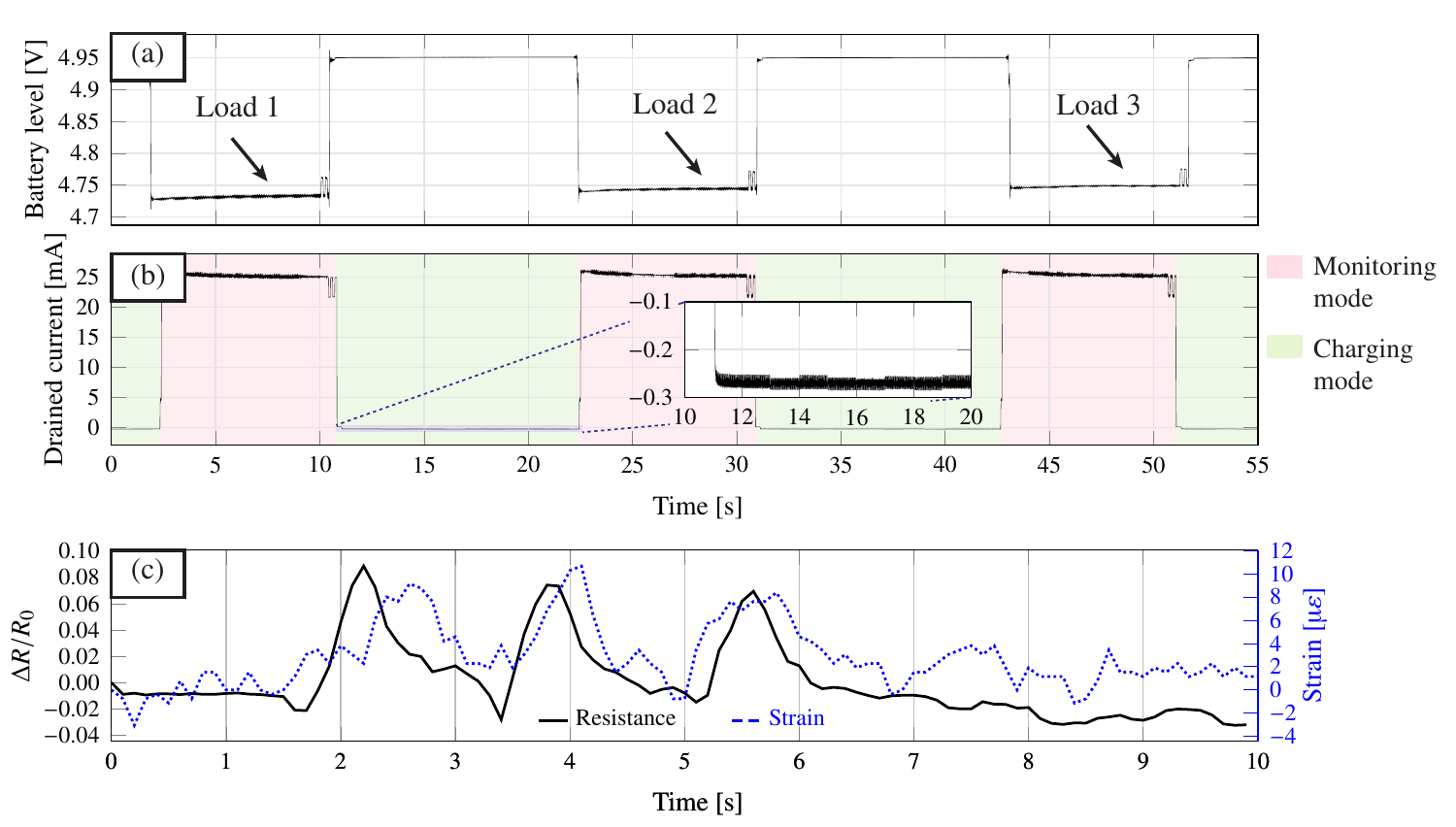}
  \caption{Simulation test of the proposed self-powered WIM system including three cycles of monitoring and charging stages. Battery voltage (a);  Electrical current time history on the outlet of the battery (+ denotes draining and - indicates charging) (b); Time series of the electrical signal acquired by the DAQ along with the strain measurements from the strain gauge during the 10-second recording corresponding to the third load step (c).}
  \label{operation}
\end{figure}

This last section presents a simulation test to appraise the integration of the DAQ/EMS and triggering systems with the self-sensing pavement. The simulation test consisted of three low compression forces exerted by the hand and separated in time at arbitrary positions on the upper surface of the composite slab. Therefore, for a total duration of the simulation test of 55 seconds, the DAQ system previously introduced in Section~\ref{SecT2} entered 3 and 4 times into the monitoring and charging modes, respectively. In this simulation test, the trigger was manually defined by a physical button grounding pin 1 of B port of the microcontroller (PB1 in Fig.~\ref{daqscheme}) when pressed. All throughout the duration of the test, the EH system with the optimal proof masses (78 g) was excited with an harmonic load with frequency coincident with the resonant frequency of the harvesters (11.3 Hz). Figures~\ref{operation} (a) and (b) report the time histories of the battery voltage and the electrical current flowing to the monitoring system. It is noted in Fig.~\ref{operation} (a) that the battery voltage experiences sudden drops around 4.75 V every time the DAQ leaves the sleep mode and starts demanding power supply. Instead, during charging periods, the battery voltage rises to around 4.95 V. Similarly, the electrical current in Fig.~\ref{operation} (b) exhibits positive values during the monitoring phase (current leaving the battery towards the DAQ) and negative values during the charging phase (power generated by the harvesters and stored in the battery). The current demand of DAQ from the battery is 25 mA during the monitoring stage, while the EH system during the charging stage is capable of providing 0.2 mA corresponding to a power level close to 1 mW. Figure~\ref{operation} (c) reports the electrical signals obtained from the slab sensor along with the measurements registered with the strain gauge for the 10 seconds recording corresponding to the third load step in Figs.~\ref{operation} (a,b). Even though the magnitude of the load was extremely low, the results in Fig.~\ref{operation} (c) evidence the high sensitivity of the slab sensor. Indeed, the strain signals exhibit remarkably inferior quality with a noticeable lower signal-to-noise ratio. This fact might be attributed to deficiencies in the attachment of the strain gauge on the sensor as well as its location with respect to the loaded surface area. Instead, the electrical output of the sensor depends upon the volumetric deformation of the slab, which allows to monitor the effect of loads applied on every position of its surface.

\section{Concluding remarks}\label{SecT4}

\rd{Built upon previous findings by the authors concerning the manufacturing, testing and field application of load-sensing composite pavements, this} work has presented an initial exploratory investigation to appraise the potential use of vibration-based EH to develop self-powered smart pavements for self-sustainable and low-cost traffic and WIM monitoring. In general, the proposed technological concept of this research lies on three main elements: piezoresistive smart pavements, vibration-based EH, and low-power DAQ/EMS. On one hand, EVIzero/CMF asphalt-like composite has been investigated as a low-cost and scalable solution to develop self-sensing pavements. Then, the potential of an EH system comprising two symmetric piezoelectric bimorph cantilevered beams connected in series to harvest traffic-induced vibrations has been theoretically and experimentally investigated. Finally, a low-power bespoke DAQ/EMS system has been developed to conduct automated WIM. The system involves two circuit units allowing to alternate between two functioning modes, that is charging and monitoring modes. The DAQ system involves a voltage reader, a micro-controller, a battery, and an USB data output unit.

The paper has presented the numerical results and discussion on an experimental campaign \rd{under laboratory conditions} conducted to assess the feasibility of the proposed technology. The experiments have comprised: (i) electromechanical characterization tests of the smart pavement and the piezoelectric cantilevered beams (with and without proof masses); (ii) assessment of the power production capability of the developed EH system; and (iii) evaluation of the performance of the complete WIM system. \rd{Overall, the reported fundamental and laboratory results provide a strong evidence of the feasibility of the presented prototype technology to be used as a low-cost self-powered WIM system, enabled by the low energy consumption of the developed smart pavement and the custom-made electronics.} The key findings of this work are summarised below:  

\begin{itemize}
	\item The characterization tests of the smart pavement sensor have \rd{verified previous findings on its self-sensing capabilities and proved the repeatability of the material production. Specifically, the presented results have} demonstrated that the sensor readings are highly correlated with the deformation of the material under compressive loads (R$^2$=0.94). In addition, the newly introduced smart pavement can operate at low energy demands (5V) and exhibits high repeatability properties, evidencing its great potential for extensive use as large-scale WIM sensors. 
	\item The DAQ prototype has proved low-power consumption of 125 mW (5V, 25 mA) and 0.03 mW (5V, 0.006 mA) during monitoring and sleep periods. Given that the DAQ system only operates in monitoring mode during 10 seconds when a vehicle is detected, the designed property can be considered ultra low-power. 
	\item The theoretical and experimental characterization tests of the energy harvesters have allowed to tune the proof masses to adapt the fundamental frequency of the piezoelectric cantilevered beams to the frequency broadband typically excited by traffic. Specifically, proof masses of 78 g decreased the fundamental frequency of the harvester to 11.3 Hz. A laboratory test replicating the vertical accelerations induced by a passing vehicle demonstrated the effectiveness of the final configuration of the piezoelectric cantilevered beams to harvest traffic-induced vibrations. 
	\item The presented experimental results have reported the designed EH system generates up to 0.1 mW under \rd{real} traffic-induced vibrations \rd{replicated in the laboratory}. Although such a power generation is rated as insufficient to sustain a self-powered design, the presented results show promise towards new improved prototypes. These may include the use of dense staking of harvesters with superior power generation performance, combination with other EH devices like solar panels, or the amplification of traffic-induced accelerations through mechanical systems.  
\end{itemize}

\rd{The presented feasibility study paves the way to the field implementation of the prototype self-powered WIM technology. In particular, future research efforts will focus on the analysis of the sensitivity to environmental (temperature and humidity) and traffic conditions, as well as to the time stability and compatibility of the developed EMS system with other complementary energy sources such as solar panels.} 

\section*{Acknowledgements}


This work is a followup study to Marie Skłodowska-Curie Innovative Training Network (MSCA ITN) \\``SAFERUP!'' project (Grant Agreement No. 765057) and was supported by Fondo integrativo speciale per la ricerca (FISR) to develop autonomous sensing systems based on smart materials.

\appendix
\section{\rd{FEM formulation of piezoceramic harvesters}}\label{appendix:FEM}

In the absence of mechanical dissipative effects, the extended Hamilton's principle applied to the bimorph piezoelectric cantilever beam model previously sketched in Fig.~\ref{Beam_model} between two time instants $t_1$ and $t_2$ reads~\cite{Meirovitch2010}:

\begin{equation}\label{FEM_beam4}
\int_{t_1}^{t_2} \left(\delta T - \delta U + \delta W_{ie} + \delta W_{nc}\right)\textrm{d}t=0,
\end{equation}

\noindent where $\delta T$, $\delta U$, $\delta W_{ie}$ and $\delta W_{nc}$ are the first variations of the total kinetic energy, the total potential energy, the internal electric energy, and the work done by non-conservative mechanical forces and electric charge components, respectively. In this formulation, the effect of base excitation is considered in the total kinetic energy term and the mechanical damping will be introduced later into the discretized equations. Hence, the only non-conservative work in Eq.~(\ref{FEM_beam4}) is due to the electric charge output $Q$ flowing to the external resistor such that $\delta W_{nc}=Q \delta V$. The total potential energy can be obtained as the sum of internal elastic strain and the electrostatic energy as:

\begin{equation}\label{FEM_beam5}
U=\frac{1}{2} \left(\int_{\Omega_s} \bm{\varepsilon}^\textrm{T}\bm{\sigma}\textrm{d}\Omega_s+\int_{\Omega_p} \bm{\varepsilon}^\textrm{T}\bm{\sigma}\textrm{d}\Omega_p \right),
\end{equation}

\noindent with subscripts $s$ and $p$ relating the corresponding magnitudes to the substructure and piezoceramic layers, respectively. The integrations in Eq.~(\ref{FEM_beam5}) are performed over the volume $\Omega$ of the respective material. 

The total kinetic energy of the beam is given by:

\begin{equation}\label{FEM_beam6}
T= \frac{1}{2}\left(\int_{\Omega_s} \rho_s \frac{\partial \tilde{\textbf{u}}^\textrm{T}}{\partial t}\frac{\partial \tilde{\textbf{u}}}{\partial t} \textrm{d}\Omega_s + \int_{\Omega_p} \rho_p \frac{\partial \tilde{\textbf{u}}^\textrm{T}}{\partial t}\frac{\partial \tilde{\textbf{u}}}{\partial t} \textrm{d}\Omega_p\right),
\end{equation}

\noindent where $\rho_s$ and $\rho_p$ denote the mass densities of the substructure and the piezoceramic layers, respectively. Vector $\tilde{\textbf{u}}$ stands for the absolute displacement vector, that is the superposition of the base displacement $W_g(t)$ and the relative displacement vector $\textbf{u}$, that is:

\begin{equation}\label{FEM_beam7}
\tilde{\textbf{u}} = \textbf{u}+\textbf{u}_b = \textbf{u}+\begin{bmatrix} 0 & 0 & w_g(t) \end{bmatrix}^\textrm{T}.
\end{equation}

The internal electrical energy in the piezoceramic layers is:

\begin{equation}\label{FEM_beam8}
W_{ie} = \frac{1}{2} \int_{\Omega_p} \textbf{E}^\textrm{T} \textbf{D} \textrm{d}\Omega_p.
\end{equation}

Under the assumption of Euler-Bernoulli beam theory (the width and thickness of the beam are very small compared to its length), the displacement field is defined as:

\begin{equation}\label{FEM_beam1}
\textbf{u} = \begin{bmatrix} u_o(x,t)-z\theta_x(x,t) & 0 & w_o(x,t) \end{bmatrix}^\textrm{T},
\end{equation}

\noindent where $\theta_x=\frac{\partial w(x,t)}{\partial x}$ denotes the rotation of the cross-sectional area around the $y$-axis, and terms $u_o(x,t)$ and $w_o(x,t)$ denote the horizontal and vertical displacements of the neutral axis. Under the assumption of linear strain-displacements, the only non-zero strain component $\varepsilon_{11}$ can be extracted as:

\begin{equation}\label{FEM_beam2}
\varepsilon_{11} = \frac{\partial u_o}{\partial x}-z\frac{\partial \theta_x}{\partial x} = \begin{bmatrix}
1 & -z
\end{bmatrix}\begin{bmatrix}
\frac{\partial u_o}{\partial x} \\
\frac{\partial \theta_x}{\partial x}
\end{bmatrix}.
\end{equation}

On this basis, the constitutive relations in Eq.~(\ref{FEM_beam3}) can be simplified for the isotropic substructure as:

\begin{equation}\label{FEM_beam3B}
\sigma_{11} = Y_s \varepsilon_{11} = Y_s \left(\frac{\partial u_o}{\partial x}-z\frac{\partial \theta_x}{\partial x}\right),
\end{equation}

\noindent and for the piezoceramic layers as:

\begin{equation}\label{FEM_beam4B}
\begin{split}
\sigma_{11} &= c_{11} \varepsilon_{11}-\bar{e}_{31}E_3 = c_{11}\left(\frac{\partial u_o}{\partial x}-z\frac{\partial \theta_x}{\partial x}\right)+\bar{e}_{31}\frac{V}{2h_p},\\
D_3 &= \bar{e}_{31}\varepsilon_{11}+\bar{\epsilon}_{33}E_3 = \bar{e}_{31}\left(\frac{\partial u_o}{\partial x}-z\frac{\partial \theta_x}{\partial x}\right)-\bar{\epsilon}_{33}\frac{V}{2h_p},
\end{split}
\end{equation}

\noindent where $Y_s$ and $c_{11}$ are the elastic moduli of the substructure layer and the piezoceramic layer at constant electric field , $\bar{e}_{31}$ is the effective piezoelectric stress constant, $\bar{\epsilon}_{33}$ is the permittivity component at constant strain, $D_3$ is the electric displacement component in the $z$-direction (i.e.~the poling direction), and $E_3$ is the electric field component. Note that in the series configuration in Fig.~\ref{Beam_model}, $\bar{e}_{31}$ has opposite signs for the top and the bottom piezoceramic layers ($\bar{e}_{31}=e_{31}$ in the top layer $\Omega_p^+$, and $\bar{e}_{31}=-e_{31}$ in the bottom layer $\Omega_p^-$). Since $\textbf{E}=-\nabla \phi$, the instantaneous electric field in Eq.~(\ref{FEM_beam4B}) is directly expressed in terms of the voltage across each piezoceramic layer as $E_3=-\frac{V}{2h_p}$. Introducing Eqs.~(\ref{FEM_beam5}), (\ref{FEM_beam6}), (\ref{FEM_beam8}) and (\ref{FEM_beam4B}) into the Hamilton's Principle in Eq.~(\ref{FEM_beam4}), one can obtain after some manipulation:

\begin{equation}\label{FEM_beam7B}
\delta T = \int_{\Omega_s} \delta \tilde{\textbf{u}} \rho_s \ddot{\tilde{\textbf{u}}} \,\textrm{d}\Omega_s + \int_{\Omega_p} \delta \tilde{\textbf{u}} \rho_p \ddot{\tilde{\textbf{u}}} \,\textrm{d}\Omega_p,
\end{equation}
\begin{equation}\label{FEM_beam8B}
\delta U = \int_{\Omega_s} \delta \varepsilon_{11} Y_s \varepsilon_{11} \textrm{d}\Omega_s + \int_{\Omega_p} \delta \varepsilon_{11} c_{11} \varepsilon_{11} \,\textrm{d}\Omega_p + \frac{1}{2}\int_{\Omega_p^+} \delta \varepsilon_{11} e_{31} E_3 \,\textrm{d} \Omega_p^+ - \frac{1}{2}\int_{\Omega_p^-} \delta \varepsilon_{11} e_{31} E_3 \,\textrm{d} \Omega_p^-,
\end{equation}
\begin{equation}\label{FEM_beam9}
\delta W_{ie} = \int_{\Omega_p} \delta E_3 \bar{\epsilon}_{33}E_3 \,\textrm{d}\Omega_p - \frac{1}{2}\int_{\Omega_p^+} \delta E_3 e_{31} \varepsilon_{11} \,\textrm{d} \Omega_p^+ + \frac{1}{2}\int_{\Omega_p^-} \delta E_3 e_{31} \varepsilon_{11} \,\textrm{d} \Omega_p^-.
\end{equation}

\noindent with overdots denoting time-derivatives. In the above variational form of the governing equations of motion, the mechanical displacement field $\textbf{u}$ and the electric potential field $V$ are the unknown functions. To solve these unknowns numerically, the governing equations are discretized with 2-nodes Euler-Bernoulli beam elements. To do so, the FEM approximation of the beam transverse deflection $w_o^e(x)$ and the horizontal displacement $u_o(x)$ in the $e$-th element are approximated by Hermitian and Lagrangian shape functions, respectively:

\begin{equation}\label{FEM_beam10}
w_o^e(x,t) = w_j^e (t)N^{\varphi}_{j}(x), \; j=1,2,3,4,
\end{equation}
\begin{equation}\label{FEM_beam11}
u_o^e(x,t) = u_j^e (t)N^{\psi}_{j}(x), \; j=1,2,
\end{equation}

\noindent where terms $w_j^e$ represent the deflection and slope at the nodes of the $e$-th element, and $u_j$ is the horizontal displacement of the $e$-th element nodes. Functions $N^{\varphi}_{j}(x)$ and $N^{\psi}_{j}(x)$ are the $j$-th Hermitian and Lagrangian shape functions, respectively. On this basis, three mechanical degrees of freedom are defined per node and organised in a vector $\textbf{d}^e$ given by:

\begin{equation}\label{FEM_beam10B}
\textbf{d}^e = \begin{bmatrix} u_1 & w_1 & \theta_1 & u_2 & w_2 & \theta_2 \end{bmatrix}^\textrm{T}.
\end{equation}

Therefore, the mechanical displacements and rotations at any $x$ coordinate in the $e$-th beam element can be expressed as:

\begin{equation}\label{FEM_beam11B}
\begin{bmatrix} u(x,t) \\
w(x,t)
\theta(x,t) \\
\end{bmatrix} = \textbf{N}_e(x)\textbf{d}^e,
\end{equation}

\noindent and the corresponding strain $\varepsilon_{11}(x,z,t)$ as:

\begin{equation}\label{FEM_beam12}
\varepsilon_{11}(x,z,t) = \begin{bmatrix}
1 & -z\\
\end{bmatrix}\textbf{B}_e(x)\textbf{d}^e,
\end{equation}

\noindent with matrices $\textbf{N}_e(x)$ and $\textbf{B}_e(x)$ being the shape function matrix and the corresponding deformation matrix, respectively. On this basis, by substituting the finite element interpolation into the extended Hamilton's principle in Eq.~(\ref{FEM_beam4}) and using the variational forms in Eqs.~(\ref{FEM_beam7B}-\ref{FEM_beam9}), the following system of ordinary differential equations is obtained for the free vibration of an arbitrary $e$-th element defined between $x$-coordinates $x_1$ and $x_2$ ($l_e=x_2-x_1$):

\begin{equation}\label{FEM_beam9B}
\begin{split}
\textbf{M}^e\ddot{\textbf{d}}^e+\textbf{K}^e\textbf{d}-\bm{\Theta}^e \, v^e(t) &= \textbf{0},\\
\left(\bm{\Theta}^e\right)^\textrm{T}\textbf{d}^e+C_p^e \, v^e(t)+Q^e=0.\\
\end{split}
\end{equation}

The element mass matrix $\textbf{M}^e$, stiffness matrix $\textbf{K}^e$, coupling matrix $\bm{\Theta}^e$, and capacitance matrix $C_p$ in Eq.~(\ref{FEM_beam9B}) are given by:

\begin{equation}\label{FEM_beam11C}
\textbf{M}^e = \int_{x_1}^{x_2} \rho_{h} \textbf{N}_e^\textrm{T}(x)\begin{bmatrix} A_{h} & 0 & 0\\
0 & A_h & 0\\
0 & 0 & I_h
\end{bmatrix} \textbf{N}_e \,\textrm{d}x,
\end{equation}
\begin{equation}\label{FEM_beam12B}
\textbf{K}^e = \int_{x_1}^{x_2} \textbf{B}_e^\textrm{T}(x) c_{11} \begin{bmatrix} A_{h} & 0 \\ 0 & I_{2,h} \end{bmatrix} \textbf{B}_e(x) \,\textrm{d}x,
\end{equation}
\begin{equation}\label{FEM_beam13}
\bm{\Theta}^e = \frac{1}{2 t_p} \int_{x_1}^{x_2} \textbf{B}_e^\textrm{T} e_{31} \begin{bmatrix} A_{p} & -I_{1,p} \end{bmatrix} \,\textrm{d}x,
\end{equation}
\begin{equation}\label{FEM_beam14}
C_p^e = \epsilon_{33} \frac{b l_e}{2h_p},
\end{equation}

\noindent with $A_{p}=bh_p$ and $I_{1,p}=A_p\left(h_s+h_p\right)/2$ being the cross-section area and the first moment of inertia of one of the piezoceramic layers. Terms $\rho_h$, $A_{h}$ and $I_h$ denote the homogenized mass density, cross-section and second moment of inertia of the composite beam:

\begin{equation}\label{FEM_beam16}
\rho_{h} = \frac{b}{A_h}\left(h_s\rho_2+2h_p\rho_p\right),
\end{equation}
\begin{equation}\label{FEM_beam17}
A_{h} = b\left(\frac{Y_s}{c_{11}}h_s+2h_p\right),
\end{equation}
\begin{equation}\label{FEM_beam18}
I_{2,h} = \frac{b}{4}\left\{\frac{Y_s}{c_{11}} h_s^3+\left[\left(h_p+\frac{h_s}{2}\right)^3-h_s^3\right]\right\}.
\end{equation}

After assembling all the $n_e$ elements in the structure, the resulting global equations of motion read:

\begin{equation}\label{FEM_beam19}
\begin{split}
\textbf{M}\ddot{\textbf{d}}+\textbf{C}\dot{\textbf{d}}+\textbf{K}\textbf{d}-\bm{\Theta} \, \textbf{v} &= \textbf{F},\\
\left(\bm{\Theta}\right)^\textrm{T}\textbf{d}+\textbf{C}_p \, \textbf{v}+\textbf{Q}=0,\\
\end{split}
\end{equation}

\noindent where $\textbf{M} \in \mathbb{R}^{n_m \times n_m}$ is the global mass matrix,  $\textbf{K} \in \mathbb{R}^{n_m \times n_m}$ is the global stiffness matrix , $\bm{\Theta} \in \mathbb{R}^{n_m \times n_e}$ is the global electromechanical coupling matrix, $\textbf{C}_p \in \mathbb{R}^{n_e \times n_e}$ is the diagonal global capacitance matrix, $\textbf{F} = -\textbf{M} \ddot{w}_g(t)\textbf{L}$ is the global vector of mechanical forces with $\textbf{L}$ the location matrix filled with ones in all the transverse translational DOFs, $\textbf{d}\in \mathbb{R}^{n_m \times 1}$ is the global vector of mechanical coordinates, and $\textbf{v}\in \mathbb{R}^{n_e \times 1}$ is the global vector of voltage outputs. Here, $n_m$ is the number of mechanical DOFs in the structure. The damping matrix $\textbf{C}$ can be easily introduced in Eq.~(\ref{FEM_beam19}) according to the proportional Rayleigh damping model, i.e.~$\textbf{C}=\alpha \textbf{M}+\beta \textbf{K}$, where proportionality coefficients $\alpha$ and $\beta$ are usually determined from experiments. 

In this formulation, the number of electrical DOFs is equal to the number of elements $n_e$. However, piezoceramics in practice come from the manufacturer with thin and very conductive electrode layers on the top and bottom surfaces. It is therefore reasonable to assume that all finite elements $n_e$ generate the same voltage output so that the elements of vector $\textbf{v}$ are identical, that is $\textbf{v}=\textbf{1}_{1 \times n_e}v_p$. In this light, the coupling and capacitance matrices are transformed as $\tilde{\bm{\Theta}} = \textbf{1}_{n_e \times 1}\bm{\Theta} \in \mathbb{R}^{n_m \times 1}$ and $C_p = \textrm{trace} \, \left( \textbf{C}_p \right)$. Furthermore, to include the effect of the electrical load, time derivation is taken from the second equation of Eq.~(\ref{FEM_beam19}). Considering that $\dot{Q}=v_p/R_l$, \rd{the global equations of motion reported in the manuscript in Eq.~(\ref{FEM_beam19B}) are obtained.}

\bibliographystyle{elsarticle-num}
\bibliography{reference}

\end{document}